\documentclass[%
notitlepage,
report,
nofootinbib,
 amsmath,amssymb,
 aps,
]{revtex4-1}

\usepackage{graphicx}
\usepackage{subfig}
\graphicspath{{/home/jripley/SphericalCollapse/EdGB_HorizonPenetrating/plots/}}
\usepackage{bm}% bold math
\usepackage{hyperref}% add hypertext capabilities
\usepackage{color}

\allowdisplaybreaks

\begin{document}

%\preprint{APS/123-QED}

\title{Scalarized Black Hole dynamics in Einstein dilaton Gauss-Bonnet Gravity}

\author{Justin L. Ripley}
\email{jripley@princeton.edu}
\author{Frans Pretorius}
\email{fpretori@princeton.edu}
% \altaffiliation[Also at ]{}%Lines break automatically or can be forced with \\
\affiliation{%
 Department of Physics, Princeton University, Princeton, New Jersey 08544, USA.
% this line break forced with \textbackslash\textbackslash
}%


\date{\today}% It is always \today, today,

\begin{abstract}
	We report on a numerical investigation of the stability
of scalarized black holes in Einstein dilaton Gauss-Bonnet (EdGB) gravity in
the full dynamical theory, though restricted to spherical symmetry.
We find evidence that for sufficiently small curvature-couplings
the resulting scalarized black hole solutions are nonlinearly 
stable.
For such small couplings, we show that an elliptic region forms inside
these EdGB black hole spacetimes (prior to any curvature singularity),
and give evidence that this region remains censored from asymptotic view.
However, for coupling values ``superextremal''
relative to a given black hole mass,
an elliptic region forms exterior to the horizon, implying the exterior
Cauchy problem is ill-posed in this regime.

%\begin{description}
%\end{description}
\end{abstract}

\maketitle

%\tableofcontents

%%%%%%%%%%%%%%%%%%%%%%%%%%%%%%%%%%%%%%%%%%%%%%%%%%%%%%%%%%%%%%%%%%%%%%%%%%%%%%
\section{Introduction}
	We present numerical results on the nonlinear evolution of spherically
symmetric black hole
solutions in a modified gravity theory:
Einstein dilaton Gauss-Bonnet (EdGB) gravity. 
EdGB gravity is one of the few known
scalar tensor theories that may admit sensible classical evolution
(at least for some open subsets of initial data; see our earlier studies
\cite{Ripley:2019hxt,Ripley:2019irj}),
yet does not allow the  
Schwarzschild or Kerr stationary black hole solutions.
Instead the expected solutions 
are conjectured to be ``scalarized'' black holes
\cite{Kanti:1995vq,Sotiriou:2013qea,Sotiriou:2014pfa}
(the detailed form of this statement depends
on the functional form of the Gauss-Bonnet coupling and
scalar field potential
\cite{PhysRevD.99.064011,Macedo:2019sem}). The variant
of EdGB gravity  
we consider is poorly constrained by weak field gravity
measurements (e.g. from binary pulsars \cite{Yagi:2015oca}), though
to be consistent with the speed of gravitational waves
inferred by the binary neutron star merger
GW170817~\cite{Abbott:2018lct} requires a negligible
cosmological background for the dilaton
field~\cite{Tattersall:2018map}\footnote{
We note that in asymptotically flat spacetimes it has been explicitly shown
that the speed of linearized tensor and scalar perturbations
in EdGB gravity approach light speed in regions
far away from gravitating sources \cite{Ayzenberg:2013wua}
}.
Assuming the latter, the strongest constraints on the theory
may then come from gravitational wave observation
of the final moments of binary black hole inspiral.
The theory thus provides an interesting alternative to general relativity
(GR) to compare against when gravity is in the strong
field dynamical regime (see e.g. \cite{Yunes:2016jcc} and references therein).

	From the perspective of effective field theory, EdGB gravity
can be motivated as the leading correction to
GR in a low energy expansion of quantum gravity
that incorporates mixing between a
scalar degree of freedom and the tensor (metric) degrees of freedom of GR 
\cite{Zwiebach:1985uq,Gross:1986mw}. Then one would not
expect significant modifications to GR away from the Planck scale, 
and in particular not for astrophysical black holes. Alternatively,
as we do here,
one could consider the coupling parameter of the theory to be 
arbitrary, and EdGB gravity taken verbatim as a classical
theory of gravity with a scale dependent modification to GR.
Such a theory may not be mathematically well-posed in some
regimes (or in a generic sense not at
all~\cite{Papallo:2017qvl,Papallo:2017ddx}), though a ``healthy''
sector of solutions could still be extracted by treating it as
an effective field theory and limiting consideration to small perturbative
corrections to GR.
Several groups are pursuing such an ``order reduction'' approach
to understanding EdGB gravity, and related theories where beyond-Ricci
curvature scalars are added to the Einstein Hilbert gravitational action~\cite{
Benkel:2016kcq,Yunes:2016jcc,Okounkova:2017yby,Okounkova:2018pql,
Witek:2018dmd,Okounkova:2019dfo,Okounkova:2019zep,Okounkova:2019zjf}. 
Another approach, inspired by the Israel-Stewart ``fix'' of relativistic 
hydrodynamics~\cite{Israel:1979wp},
is to explicitly modify the GR-extensions to 
lead to well-posed equations~\cite{Cayuso:2017iqc,Allwright:2018rut}.

Our approach is instead to attempt to solve the complete
classical field theory, and discover which classes of initial data (if any)
lead to well-posed evolution. Our motivation is two fold.
The first is the desire to know how classical gravity can in principle differ
from the predictions of GR in the dynamical strong field regime,
as is applicable
to the last stages of binary black hole coalescence. This could give more
meaning to quantitative statements of consistency of observed waveforms
with the predictions of GR, help constrain EdGB gravity, or discover
modifications to GR of a class similar to that offered by EdGB gravity.
Though an effective field theory approach as described above is likely 
``guaranteed'', by construction, to give well-posed evolution schemes for small
deviations to GR, it is still unknown if this approach could be pushed
to solve for modifications large enough to provide waveforms distinguishable
from GR in an observation, given the typical signal-to-noise ratios
expected from the current generation of ground based detectors,
and taking waveform degeneracies into account.

The second reason is that nonlinear modifications of
gravity have been introduced in attempts to address the discovery of
dark matter, dark energy, solve the flatness and horizon problems
of early universe cosmology, and resolve the issues of the initial
cosmological singularity and singularities formed during
gravitational collapse. EdGB is an important and representative member
of a class of modified gravity theories, Horndeski theories,
that have been invoked to solve these various issues within
classical GR (for a recent review on Horndeski
theories see \cite{Kobayashi:2019hrl}),
and it thus is interesting to see if the theory is mathematically
sensible as a classical field theory.

	Previous studies of EdGB black holes
have focused on static
solutions to the theory
(e.g. \cite{Sotiriou:2014pfa,PhysRevD.99.064011,Macedo:2019sem}),
the dynamics of the EdGB scalar in the
\emph{decoupling limit}
(e.g. \cite{Benkel:2016kcq,Benkel:2016rlz,Witek:2018dmd}),
or linear perturbations of static EdGB black hole
backgrounds (e.g. \cite{Blazquez-Salcedo:2016enn}).
Additionally, a recent work \cite{Okounkova:2019zep}
explored the dynamics of the scalar
and metric fields of the theory
from an effective-field theory framework.
There it was shown that scalarized, rotating
EdGB black hole
solutions are stable for small enough couplings, up to leading
order in metric and scalar field perturbations.
Restricted to spherical symmetry, our results extend this to
all orders in the Gauss-Bonnet coupling, showing consistency
for small couplings, and showing where and how the theory
breaks down for large couplings.

In a previous work, we studied the dynamics of EdGB gravity
in spherically symmetric collapse using horizon-avoiding coordinates
\cite{Ripley:2019hxt,Ripley:2019irj}. There we considered
collapse of a dense concentration of the dilaton field,
and for sufficient mass could show a horizon was forming, though
we could not evolve beyond that time.
Here, we give results from a new code solving the EdGB equations
in horizon penetrating coordinates, allowing us to investigate
the long-term stability of scalarized black holes
(for times up to of order $t\sim 10^3 m$, where $m$ is the mass
of the black hole).
Also, we begin with the Schwarzschild
solution as initial data (with an optional exterior dilaton field
perturbation).

Upon evolution of 
Schwarzschild initial data, we find that the scalar hair grows, and an elliptic
region forms in the interior of the black hole. 
This indicates black hole physics in EdGB gravity has aspects
of it governed by a mixed elliptic-hyperbolic equation
(or simply {\em mixed-type} equation), 
and it is unclear
how this could affect the Cauchy problem exterior to the horizon.
I.e., there is no {\em a-priori} reason to expect this elliptic
region to ``obey'' cosmic censorship, and leave the scalar in the exterior
domain to be governed by a hyperbolic partial differential equation (PDE). 
Instead, we will simply {\em assume} that this is possible, and {\em excise}
the elliptic region from the domain. If during subsequent evolution
no new elliptic region forms, and the solution settles to a stationary
state, we will claim this is a self-consistent application of excision,
and the resulting hairy black hole is stable (to within limitations
of numerical evolution)\footnote{Also in this case, that we
can freely specify the initial data for all characteristics
is not in contradiction with the result of Morawetz on
the Tricomi mixed-type equation~\cite{dirichlet_tricomi_morawetz},
which seems the relevant mixed-type equation for EdGB
gravity here~\cite{Ripley:2019irj};
rather, following the excision philosophy, we simply
do not care what irregularities or lack of uniqueness 
occur in the interior of the excised region.}. 
We do find this to be the case
for Gauss-Bonnet couplings below an extremal limit for a given black hole mass.
We compare these solutions to the scalarized decoupled
black hole solutions of EdGB gravity, and find good agreement, the better
the smaller the Gauss-Bonnet coupling is (for a fixed black hole mass).
However, above the extremal limit, an elliptic region
does form outside the horizon, indicating a break-down of the exterior
Cauchy problem
for small black holes (relative to the EdGB coupling scale).

An outline of the rest of the paper is as follows. In Sec.~\ref{sec:eom}
we describe the equations of motion, variables, and metric ansatz we use.
In Sec.~\ref{sec:code} we describe aspects of the numerical code,
including our excision strategy, as well as some diagnostic quantities
we monitor. In Sec.\ref{sec:results} we describe the results mentioned
above in detail, and end in Sec.~\ref{sec:discussion} with concluding remarks.
We leave some convergence results, and a derivation of the decoupling
limit about a Schwarzschild black hole in Painlev\'{e}-Gullstrand coordinates,
to the appendices.
	We use geometric units ($8\pi G=1$, $c=1$)
and follow the conventions of
Misner, Thorne, and Wheeler \cite{misner1973gravitation}. 
%%%%%%%%%%%%%%%%%%%%%%%%%%%%%%%%%%%%%%%%%%%%%%%%%%%%%%%%%%%%%%%%%%%%%%%%%%%%%%
\section{Equations of motion}
\label{sec:eom}
	The action for the EdGB gravity theory we consider is 
\begin{equation}
\label{eq:EdGBAction}
	S = \frac{1}{2}\int d^4x\sqrt{-g}
	\left(
		R - (\nabla\phi)^2 + 2\lambda f(\phi)\mathcal{G}
	\right)
	,
\end{equation}
	where $f(\phi)$ is a (so far unspecified) function,
and $\mathcal{G}$ is the Gauss-Bonnet scalar
\begin{equation}
\label{eq:GaussBonnetDefinition}
	\mathcal{G}
	\equiv
	\frac{1}{4}\delta^{\mu\nu\alpha\beta}_{\rho\sigma\gamma\delta}
	R^{\rho\sigma}{}_{\mu\nu}R^{\gamma\delta}{}_{\alpha\beta}
	,
\end{equation}
	where $\delta^{\mu\nu\alpha\beta}_{\rho\sigma\gamma\delta}$ is the
generalized Kronecker delta. In geometric units,
the Gauss-Bonnet coupling constant $\lambda$
has dimensions length squared. 
Varying \eqref{eq:EdGBAction} with respect to the metric and
scalar fields, the EdGB equations of motion are 
\begin{subequations}\label{eqns:einstein_eqns}
\begin{eqnarray}
\label{eq:tensor_eom}
	E^{(g)}_{\mu\nu}&\equiv&
	R_{\mu\nu} - \frac{1}{2}g_{\mu\nu}R
	+ 2\lambda\delta^{\gamma\delta\kappa\lambda}_{\alpha\beta\rho\sigma}
	R^{\rho\sigma}{}_{\kappa\lambda}
	\left(\nabla^{\alpha}\nabla_{\gamma}f(\phi)\right)
	\delta^{\beta}{}_{(\mu}g_{\nu)\delta}
	- T_{\mu\nu} 
	= 0 
	, \\
	T_{\mu\nu}
 	&=&
	\nabla_{\mu}\phi\nabla_{\nu}\phi
-	\frac{1}{2}g_{\mu\nu}(\nabla\phi)^2
	, \nonumber \\
\label{eq:scalar_eom}
	E^{(\phi)}&\equiv&
		\nabla_{\mu}\nabla^{\mu}\phi 
	+ 	\lambda f^{\prime}(\phi)\mathcal{G} 
	= 0
	.
\end{eqnarray}
\end{subequations}
	In this work we will only consider the coupling function
\begin{align}
	f(\phi)
	=
	\phi
	.
\end{align} 
	While other coupling functions are often considered in the literature
on EdGB black holes, this is the simplest which is thought to give
rise to stable scalarized black hole solution;
see
\cite{Kanti:1995vq,Berti_2015,
Sotiriou:2014pfa,Benkel:2016kcq,Witek:2018dmd} and references therein.
This coupling may additionally be motivated as
the lowest order term in the effective field theory expansion of a metric theory
coupled to a scalar field
(e.g. \cite{Yagi:2015oca}). From the symmetry
$\lambda\to-\lambda,\phi\to-\phi$, we only consider $\lambda\geq0$.

	We evolve this system in Painlev\'{e}-Gullstrand (PG)-like coordinates 
        (e.g. \cite{Adler:2005vn,Ziprick:2008cy,Kanai:2010ae,Ripley:2019tzx})
\begin{align}
	ds^2
	=
-	\alpha(t,r)^2dt^2
+	\left(dr+\alpha(t,r)\zeta(t,r)dt\right)^2
+	r^2\left(
		d\vartheta^2
	+	\mathrm{sin}^2\vartheta d\varphi^2
	\right)
	,
\end{align} 
so-named since $t={\rm const.}$ cross sections are spatially flat 
(the Schwarzschild black hole in these coordinates
is given by $\alpha=1,\zeta=\sqrt{2 m/r}$).
	
	We define the variables
\begin{subequations}
\label{eqns:defs_PQ}
\begin{eqnarray}
\label{eq:def_Q}
	Q
	\equiv
	\partial_r\phi
	, \\
\label{eq:def_P}
	P
	\equiv
	\frac{1}{\alpha}\partial_t\phi
-	\zeta Q	
	,
\end{eqnarray}
\end{subequations}
	and take algebraic combinations of Eq.~\eqref{eq:scalar_eom} and
the $tr$, $rr$, and $\vartheta\vartheta$ components of
Eq.~\eqref{eq:tensor_eom}
(c.f. \cite{Ripley:2019irj}) to obtain the following evolution equation
for the $\{P,Q\}$ variables:
\begin{subequations}
\label{eqns:evolution_eqns}
\begin{eqnarray}
\label{eq:evolution_Q}
	E_{(Q)}\equiv
	\partial_tQ
-	\partial_r\left(
		\alpha\left[
			P
		+	\zeta Q
		\right]
	\right)
	=
	0
	, \\
\label{eq:evolution_P}
	E_{(P)}\equiv
	\mathcal{A}_{(P)}\partial_tP
+	\mathcal{F}_{(P)}
	=
	0
	. 
\end{eqnarray}
\end{subequations}
	The quantities 
$\mathcal{A}_{(P)}$ and $\mathcal{F}_{(P)}$ are lengthy
expressions of 
$\{\alpha,\zeta,P,Q\}$ and their radial derivatives. We present their
explicit forms in Appendix \ref{appendix:form_of_long_terms}.
In the limit $\lambda=0$ Eq.~\eqref{eq:evolution_P} reduces to
\begin{align}
	\partial_tP
-	\frac{1}{r^2}\partial_r\left(r^2\alpha\left[Q+\zeta P\right]\right)
	=
	0
	.
\end{align} 
Interestingly, in PG coordinates the
Hamiltonian and momentum constraints do not
change their character as elliptic differential equations
going from GR to EdGB gravity:  
\begin{subequations}
\label{eqns:constraint_eqns}
\begin{eqnarray}
\label{eq:Hamiltonian_constraint}
	E^{(g)}_{\mu\nu}n^{\mu}n^{\nu}
	&\propto& % \nonumber \\
	\partial_r\left(
		\left(r - 8\lambda f^{\prime}Q\right)\alpha^2\zeta^2
	\right)
	-	8\lambda f^{\prime}\frac{P}{\alpha}
		\partial_r\left(\alpha^3\zeta^3\right)
	-	r^2\alpha^2\rho
	= 0
	, \\
\label{eq:momentum_constraint}
	E^{(g)}_{\mu r}n^{\mu}
	&\propto& % \nonumber \\
	\left(
		1
	-	8\lambda f^{\prime}\frac{\zeta}{r}P
	-	8\lambda f^{\prime}\frac{Q}{r}
	\right)\zeta
		\partial_r\alpha
	-	\frac{1}{2}r\alpha j_r \nonumber \\
	&\ & +	2\lambda f^{\prime}\frac{Q}{r\alpha^2}
			\partial_r\left(\alpha^2\zeta^2\right)
	+	4\lambda \frac{\zeta}{r}\partial_r\left(f^{\prime}P\right)
	= 0
	,
\end{eqnarray}
\end{subequations}
	where 
\begin{subequations}
\label{eqns:defs_source_terms}
\begin{eqnarray}
\label{eq:def_rho}
	\rho
	&\equiv&
	n^{\mu}n^{\nu}T_{\mu\nu}
	=
	\frac{1}{2}\left(P^2+Q^2\right)
	, \\
\label{eq:def_jr}
	j_{r}
	&\equiv&
-	\gamma_{r}{}^{\mu}n^{\nu}T_{\mu\nu}
	=
	-PQ
	,
\end{eqnarray}
\end{subequations}
	$f^{\prime}\equiv df/d\phi$,
and $n_{\mu}\equiv(-\alpha,0,0,0)$. While
Eqs.~\eqref{eq:Hamiltonian_constraint} and \eqref{eq:momentum_constraint} 
hold for any $f$, as mentioned above
we only consider $f(\phi)=\phi$ in this article.
%%%%%%%%%%%%%%%%%%%%%%%%%%%%%%%%%%%%%%%%%%%%%%%%%%%%%%%%%%%%%%%%%%%%%%%%%%%%%%
\section{Description of code and simulations}
\label{sec:code}

%%%%%%%%%%%%%%%%%%%%%%%%%%%%%%%%%%%%%%%%%%%%%%%%%%%%%%%%%%%%%%%%%%%%%%%%%%%%%%
\subsection{Diagnostics}
\label{subsec:diagnostics}
	As PG coordinates
are spatially flat the Arnowitt-Deser-Misner mass
prescription always evaluates to zero,
and does not capture the correct physical mass of the spacetime.
Instead then we use the Misner-Sharp mass \cite{PhysRev.136.B571} 
\begin{align}
	m_{MS}(t,r) 
	= 
	\frac{r}{2}\left(1-(\nabla r)^2\right)
	=	
	\frac{r}{2}\zeta(t,r)^2
	.
\end{align} 
evaluated at spatial infinity
to define the spacetime mass
(and this does give the correct mass for the Schwarzschild solution in GR)
\begin{align}
	m\equiv \lim_{r\to \infty}m_{MS}(t,r).
\end{align}
The Misner-Sharp mass can be thought of as the charge associated with the 
Kodama current, which is conserved in any spherically symmetric spacetime
(regardless of whether the Einstein equations hold )
\cite{Kodama:1979vn,Maeda:2007uu,Abreu:2010ru}.
Going through this ``derivation'' of the Misner-Sharp mass, we find that
we can think of $m_{MS}(t,r)$ as representing the radially integrated
energy density of the following (conserved) stress-energy tensor for EdGB
gravity \cite{Ripley:2019irj}
\begin{align}
\label{eq:stress_energy_EdGB}
	\mathcal{T}_{\mu\nu}
	\equiv&
-	2\lambda
	\delta^{\gamma\delta\kappa\lambda}_{\alpha\beta\rho\sigma}
	R^{\rho\sigma}{}_{\kappa\lambda}
	\left(\nabla^{\alpha}\nabla_{\gamma}\phi\right)
	\delta^{\beta}_{(\mu}g_{\nu)\delta}
+	\nabla_{\mu}\nabla_{\nu}\phi
-	\frac{1}{2}g_{\mu\nu}\left(\nabla\phi\right)^2
	.
\end{align}
	There is no other definition for a covariantly conserved
stress-energy tensor
that does not involve both the Riemann tensor and derivatives of the scalar
field (besides the Einstein tensor itself,
or a constant times the stress-energy tensor)
in EdGB gravity.
We note that the stress-energy tensor \ref{eq:stress_energy_EdGB} also
conforms with earlier choices for the stress-energy
tensor of EdGB gravity \cite{Kanti:1995vq}.

As described in the next section we compactify so that infinity is
at a finite location on our computational grid. The asymptotic mass $m$
is preserved up to truncation error in our simulations of both
EdGB gravity and GR.
Given a spacetime with mass $m$, we define the dimensionless 
curvature-coupling
\begin{equation}\label{c_coup}
	C
	\equiv
	\frac{\lambda}{m^2}.
\end{equation}
	We will classify different
solutions based on their curvature couplings $C$, with
GR the limit $C=0$;
empirically (as we discuss in our results \ref{sec:results})
we find strong EdGB corrections arising when $C\gtrsim 0.1$.

	Following the procedure used in
\cite{Ripley:2019hxt,Ripley:2019irj}, we calculate
the radial characteristics of the scalar degree of freedom
via Eqs.~\eqref{eq:evolution_Q} and \eqref{eq:evolution_P},
after having removed the spatial derivatives of $\alpha$ and $\zeta$
from these equations using the constraints (\ref{eq:Hamiltonian_constraint},
\ref{eq:momentum_constraint}).
The corresponding characteristic speeds $c_{\pm}$ are
\begin{align}
	c_{\pm}\equiv\mp\xi_t/\xi_r,
\end{align}
where
$\xi_a\equiv(\xi_t,\xi_r)$ 
solves the characteristic equation
\begin{align}
\label{eq:characteristic_equation}
	\mathrm{det}\left[
		\begin{pmatrix}
			\delta E_{(P)}/\delta(\partial_aP)
		&	\delta E_{(P)}/\delta(\partial_aQ)
		\\	\delta E_{(Q)}/\delta(\partial_aP)
		&	\delta E_{(Q)}/\delta(\partial_aQ)
		\end{pmatrix}
		\xi_a
	\right]
	=
	0
	.
\end{align}
In the limit $\lambda=0$, these speeds reduce to the radial null
characteristic speeds in PG coordinates $c^{(n)}_\pm$ 
\begin{align}
\label{eq:null_radial_characteristics}	
	c^{(n)}_{\pm}
	\equiv
	\alpha\left(\pm1-\zeta\right)
	.
\end{align} 
	We see that $\zeta=1$ marks the location of a
marginally outer trapped surface (MOTS)
(e.g. \cite{Thornburg2007} and references therein).
We take the location of the MOTS
to represent the size of the black hole on any given time slice.

	The characteristic equation, Eq.~\eqref{eq:characteristic_equation},
takes the following form when expressed as an equation for the characteristic
speeds $c$
\begin{align}
\label{eq:characteristic_equation_quadratic}
	\mathcal{A}c^2+\mathcal{B}c+\mathcal{C}=0
	,
\end{align}
	where
\begin{subequations}
\begin{align}
	\mathcal{A}
	\equiv &
	\frac{\delta E_{(P)}}{\delta\left(\partial_tP\right)}
	\frac{\delta E_{(Q)}}{\delta\left(\partial_tQ\right)}
-	\frac{\delta E_{(P)}}{\delta\left(\partial_tQ\right)}
	\frac{\delta E_{(Q)}}{\delta\left(\partial_tP\right)}
	, \\
	\mathcal{B}
	\equiv & 
-	\left(
	\frac{\delta E_{(P)}}{\delta\left(\partial_tP\right)}
	\frac{\delta E_{(Q)}}{\delta\left(\partial_rQ\right)}
-	\frac{\delta E_{(P)}}{\delta\left(\partial_tQ\right)}
	\frac{\delta E_{(Q)}}{\delta\left(\partial_rP\right)}
	\right)
-	\left(
	\frac{\delta E_{(P)}}{\delta\left(\partial_rP\right)}
	\frac{\delta E_{(Q)}}{\delta\left(\partial_tQ\right)}
-	\frac{\delta E_{(P)}}{\delta\left(\partial_rQ\right)}
	\frac{\delta E_{(Q)}}{\delta\left(\partial_tP\right)}	
	\right)
	, \\
	\mathcal{C}
	\equiv &
	\frac{\delta E_{(P)}}{\delta\left(\partial_rP\right)}
	\frac{\delta E_{(Q)}}{\delta\left(\partial_rQ\right)}
-	\frac{\delta E_{(P)}}{\delta\left(\partial_rQ\right)}
	\frac{\delta E_{(Q)}}{\delta\left(\partial_rP\right)}
	.
\end{align}
\end{subequations}
Where the discriminant
$\mathcal{D}\equiv\mathcal{B}^2-4\mathcal{A}\mathcal{C}>0$
are regions of spacetime where the equations are hyperbolic,
where $\mathcal{D}<0$ the equations are elliptic,
and following the language of mixed-type PDEs
(e.g. \cite{otway2015elliptic} and references therein),
the co-dimension one surfaces where $\mathcal{D}=0$ separating elliptic
and hyperbolic regions are called sonic lines.
 In the GR limit $\lambda=0$ the scalar equations are always
 hyperbolic ($\mathcal{D}>0$), though as we found 
 in \cite{Ripley:2019hxt,Ripley:2019irj}, for sufficiently
 strong couplings $C$ the discriminant $\mathcal{D}$ is not
 of definite sign, and the scalar equations are then of mixed-type 
(similar conclusions have been drawn for other member of the
Horndeski class of theories;
see e.g.~\cite{Leonard:2011ce,Akhoury:2011hr,Bernard:2019fjb}).
As we are working in spherical symmetry, the tensor degrees of
freedom are pure gauge.
Our hyperbolicity analysis can thus be thought
of as applying to the scalar ``sector'' of EdGB gravity.
\footnote{Note that in a less symmetrical spacetime more
care would need to be taken to
distinguish between scalar and tensor dynamics due to the derivative
coupling between the scalar and metric fields in the EdGB
equations of motion; see Eqs.~\ref{eqns:einstein_eqns}.
}

%%%%%%%%%%%%%%%%%%%%%%%%%%%%%%%%%%%%%%%%%%%%%%%%%%%%%%%%%%%%%%%%%%%%%%%%%%%%%%
\subsection{Spatial compactification}
\label{subsec:spatial_compactification}
In vacuum when $P=Q=0$, the
general solution to Eqs.~\eqref{eq:Hamiltonian_constraint} and
\eqref{eq:momentum_constraint} is $\zeta\propto r^{-1/2}$ and $\alpha=const.$ 
\footnote
{In EdGB gravity curvature always sources a scalar field, though
for $r\gg m$ for an isolated source in an asymptotically flat spacetime,
the fall off of the curvature-sourced scalar field is sufficiently
fast not to alter, through back reaction,
the fall off of the metric derived when $P=Q=0$.}.
We found that this falloff in $\zeta$ made
it difficult to impose
stable outer boundary conditions at a fixed, finite $r$.
To alleviate this problem,
we spatially compactify through a stereographic projection
\begin{equation}
\label{eq:spatial_compactification}
	r \equiv \frac{x}{1-x/L}
	,
\end{equation}
	where $L$ is a constant, and discretize along a uniform
grid in $x$, with spatial infinity $x=L$ now the outer boundary
of our computational domain. For all the simulations presented
in this article we chose $L=5 m$, where $m$ is the mass of
the initial Schwarzschild black hole.
At $x=L$ we impose the conditions
$	\alpha |_{x=L}
	= 
	1
$ ,
$	\zeta |_{x=L}
	= 
	0 
$,
$	P |_{x=L}
	= 
	0
$,
$	Q |_{x=L}
	= 
	0
$,
$	\phi |_{x=L}
	=  
	0
$.
	These conditions are consistent with our initial conditions 
and asymptotic fall off of the metric
and scalar field. For the latter,
generally $\phi \rightarrow 1/r$, though 
if we impose exact Schwarzschild initial data outside some radius
$r_1$ (such that $\phi(r>r_1,t=0) =0$ and $\partial_t \phi(r>r_1,t=0)=0$),
the Gauss-Bonnet curvature will source an asymptotic field that
decays like $1/r^4$; by causality (as long as the equations are hyperbolic)
the $1/r$ component sourced by the black hole, or any scalar radiation from
a field we put in at $r<r_1$, will never reach spatial infinity.
%%%%%%%%%%%%%%%%%%%%%%%%%%%%%%%%%%%%%%%%%%%%%%%%%%%%%%%%%%%%%%%%%%%%%%%%%%%%%%
\subsection{Initial data}
	The computational domain covers $x\in[x_{exc},L]$
($r\in[r_{exc},\infty]$), where $x_{exc}$ ($r_{exc}$) is the excision
radius, and can vary with time (described in the following section).
We set initial data at $t=0$ by 
specifying the values of $P$ and $Q$, and then solve for $\alpha$ and $\zeta$
using the momentum and Hamiltonian constraints.
These ordinary differential equations (ODEs) are discretized using
the trapezoid rule and solved with a Newton relaxation method,
integrating from $x=x_{exc}$ to $x=L$. At $x=x_{exc}$ (some distance inside
the horizon, as discussed in the next section) we set $\alpha$ and $\zeta$
to their Schwarzschild values:
\begin{equation}
\label{eq:al_ze_id}
	\alpha|_{t=0,x=x_{exc}}=1, \qquad
	\zeta|_{t=0,x=x_{exc}}=\sqrt{\frac{2m}{r(x_{exc})}}.
\end{equation}
If we begin with
zero scalar field energy ($Q|_{t=0}=P|_{t=0}=0$), solving
the constraints recovers the Schwarzschild solution on $t=0$ to
within truncation error. 
%%%%%%%%%%%%%%%%%%%%%%%%%%%%%%%%%%%%%%%%%%%%%%%%%%%%%%%%%%%%%%%%%%%%%%%%%%%%%%
\subsection{Excision} 
\label{subsec:evolution}
At every time
step we solve for $\alpha$, $\zeta$, $P$, and $Q$ by alternating between
an iterative Crank-Nicolson solver for $P$ and $Q$ and the ODE solvers for
$\alpha$ and $\zeta$, until the discrete infinity norm
of all the residuals are below a pre-defined tolerance (typically
the tolerance was $\lesssim10^{-10}$, smaller than the typically size of
the one-norm of the independent residuals. This strategy is a 
similar strategy to that used in our earlier code based on Schwarzschild-like
coordinates, and more details can be found in~\cite{Ripley:2019irj}).
The excision strategy assumes all characteristics of hyperbolic equations
are pointing out of the domain at the excision surface $x=x_{exc}$. This
implies that for $P$ and $Q$ we cannot set boundary conditions there,
rather their evolution equations must be solved, with the finite
difference stencils for the radial derivatives appropriately changed
to one-sided differences. For $\alpha$,
as with the initial data, the inner boundary condition 
is arbitrary, and after each iteration we rescale it so that 
$\alpha(t,x=L)=1$.  
For $\zeta$, to
obtain a consistent solution to the full field equations requires that the
boundary condition $\zeta(t,x=x_{exc})$ be set by solving
the corresponding evolution equation for $\zeta$ there
(optionally $\zeta$ could be evolved over the entire domain
using this equation).
Taking algebraic combinations of the equations of motion (\ref{eq:tensor_eom}), 
an appropriate evolution equation for $\zeta$ can be obtained
\begin{align}
\label{eq:free_evolution_zeta}
	\mathcal{A}_{(\zeta)}\partial_t\zeta
+	\mathcal{F}_{(\zeta)}
	=
	0
	,
\end{align}
	The expressions for
$\mathcal{A}_{(P)}$ and $\mathcal{F}_{(P)}$ are lengthy
expressions of $\lambda$,
$\{\alpha,\zeta,P,Q\}$, and their radial derivatives. We defer
showing their full form to Appendix \ref{appendix:form_of_long_terms}. 
In the limit $\lambda=0$, Eq.~\eqref{eq:free_evolution_zeta} reduces to
\begin{align}
	\partial_t\zeta
-	\alpha\zeta\partial_r\zeta
-	\frac{\alpha}{2r}\zeta^2
-	\frac{r}{2\zeta}T_{tr}
	=
	0
	.
\end{align}
Eq.~\eqref{eq:free_evolution_zeta} provides
the boundary condition for $\zeta$ at the excision
surface. We then integrate outwards in $r$
using the Hamiltonian constraint,
Eq.~\eqref{eq:Hamiltonian_constraint}
as described above
to solve for $\zeta$. 

The formation of a MOTS is signaled by
$\zeta=1$ (Eq.~\eqref{eq:null_radial_characteristics}), and we always
place the excision surface inside the MOTS\footnote{In all cases we have
considered we find that
the ``characteristic horizon'' (the location where $c_+<0$ for the EdGB
scalar field) is exterior to
the MOTS, so placing the excision point interior to the MOTS should lead to
well posed evolution, provided the equations of motion for the EdGB
scalar remain hyperbolic.}. The location of $x_{exc}$ is updated every
time step before solving for the scalar and metric fields. The location of
the excision point is chosen so that it is always interior to the MOTS,
but lies exterior to (or directly on) the sonic line
(for further discussion see Sec.~\ref{subsec:diagnostics}
and Sec.~\ref{subsec:internal_structure_elliptic}). The location
for $r_{exc}$ on our initial data slice depended on the strength
of the curvature coupling $C$, which we detail in our Results,
Sec.~\ref{sec:results}.
%%%%%%%%%%%%%%%%%%%%%%%%%%%%%%%%%%%%%%%%%%%%%%%%%%%%%%%%%%%%%%%%%%%%%%%%%%%%%%

\subsection{Fixed mesh refinement with a hyperbolic-ODE system}
\label{subsec:fixed_mesh_refinement}
	To achieve the necessary long term accuracy over
thousands of $m$ in evolution using limited computational resources,
we evolved some simulations using a Berger and Oliger (BO)
style mesh refinement algorithm~\cite{1984JCoPh..53..484B}.
Due to the nature of our initial data and perturbations,
a fixed hierarchy suffices,
with the higher resolution meshes confined to smaller volumes
centered about the origin. For those runs, we typically used 4 additional
levels beyond the base (coarsest) level,
with a $2:1$ refinement ratio between levels;
specifically, we set the inner boundary for all levels at $x_{exc}$, and
the outer refinement boundary locations at
$x_l/m=\{5.00,2.30,2.00,1.75,1.55\}$, from coarsest to finest
(the initial horizon location $r=2m$, and we chose the
excision radius $x_{exc}$ so that $r_{exc}(x_{exc})=0.8\times 2m$;
see Eq.~\eqref{eq:spatial_compactification}).

The original BO algorithm was designed for purely
hyperbolic systems of equations; to include the ODE constraint equations,
we employ the \emph{extrapolation and
delayed solution} modification developed for such coupled elliptic/hyperbolic 
systems~\cite{Pretorius:2005ua}. 
Here, for the hyperbolic equations (governing $P$ and $Q$),
the solution is obtained
on the mesh hierarchy with the usual BO time-stepping procedure :
one time step is first taken on a coarse {\em parent} level before
two\footnote{because of our $2:1$ refinement ratio in space and time.} steps
are taken on the next finer {\em child} level, and this is repeated
recursively down the mesh hierarchy. During this phase
the ODEs are not solved, and where the values
of the corresponding constrained variables ($\alpha$ and $\zeta$)
are needed to evaluate terms in the hyperbolic equations, approximations
for these variables
are obtained via extrapolation from earlier
time levels.
Instead, the ODEs are solved after the fine-to-coarse level 
injection phase of the hyperbolic variables, when the advanced
time of a given parent level is in sync with all overlapping 
child levels (thus, on the very finest level 
this scheme reduces to the unigrid algorithm 
described in the previous section).
For more details see~\cite{Pretorius:2005ua}.

\begin{figure*}
\includegraphics[width=.8\linewidth]{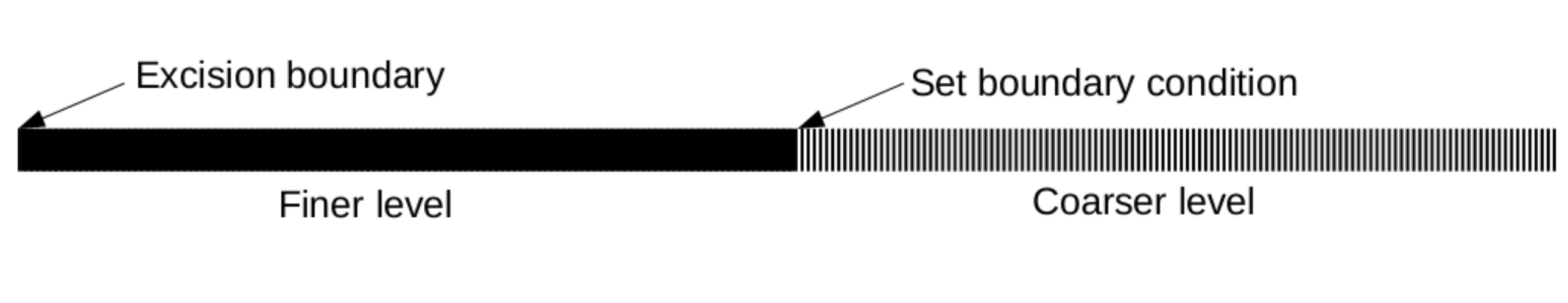}%
\caption{
Schematic illustration of the solution of ODEs at synchronized levels
with fixed mesh refinement in our setup: we integrate from left
to right (the compactified radial coordinate $x$ increases from left to right).
}
\label{fig:mesh_refinement_diagram}
\end{figure*}

One difference with our system of equations compared to that
described in~\cite{Pretorius:2005ua}, is there some
form of global relaxation method was assumed for the elliptics,
while here the ODE nature of our constraint equations requires integration
from the inner to outer boundary. This might complicate things
for a general hierarchy with disconnect grids on a given level.
Here, since we only have one grid per level, and each 
always includes the physical inner boundary, it is reasonably
straight forward to integrate the ODEs during the solution
phase of the algorithm (see also section 6.7 of \cite{Choptuik_1986}):
we begin on the finest level,
setting the boundary conditions as required at $x=x_{exc}$, then
integrate outward, using the solution at the last
point on a child level as an initial condition for continued
integration on the parent level (with the solution at interior points
on the child level injected to common points on the parent level).
This is schematically shown
for two levels in Fig.~\ref{fig:mesh_refinement_diagram}.
For $\zeta$, the boundary condition at $x=x_{exc}$ is obtained
by its corresponding evolution equation as described in the previous
section; 
$\alpha$ at $x=x_{exc}$ is set by extrapolation
from prior time levels, and we only globally rescale $\alpha$
to satisfy our outer boundary condition $\alpha(t,x=L)=1$ at times
when all levels of the hierarchy are in sync. 

%%%%%%%%%%%%%%%%%%%%%%%%%%%%%%%%%%%%%%%%%%%%%%%%%%%%%%%%%%%%%%%%%%%%%%%%%%%%%%
\section{Scalarized black holes: numerical results}
\label{sec:results}
%%%%%%%%%%%%%%%%%%%%%%%%%%%%%%%%%%%%%%%%%%%%%%%%%%%%%%%%%%%%%%%%%%%%%%%%%%%%%%
\subsection{Overview of simulations and plots}
\label{sec:overview_plots}
	To help keep track of the various simulation results we
present, we collect some of our simulation parameters in 
Table.~\eqref{table:fig_details}.
We found for long time evolution simulations ($t\gtrsim10^3m$) fixed mesh
refinement was essential to maintain high accuracy evolution
($\delta m/m\lesssim10^{-2}$). Unigrid evolution was sufficient for shorter
simulation runs.
When we quote a value of resolution $\Delta x$, it represents
the resolution of the coarsest (base) level if fixed mesh refinement
was used, otherwise it is the resolution of the unigrid mesh.
We found that stably resolving the initial growth of the sonic
line that formed inside the EdGB black hole required using
smaller Courant-Friedrichs-Lewy (CFL) numbers,
but CFL numbers as large as $0.5$ led to stable
evolution if we excised well away from the sonic line.
\begin{table}
\begin{tabular}{ c|c|c|c|c } 
 \hline
  Figure & grid layout & grid resolution $\Delta x$ 
& CFL number & initial data\\ 
 \hline
 2,3,4,5,6,12 & fixed mesh refinement & $0.38$ & $0.25$ & Schwarzschild \\ 
 7  & unigrid & $0.024$ & $0.5$ & Schwarzschild with bump:\\ 
    &         &         &       & $\phi_0=0.1$,$a=45$,$b=55$. \\ 
 8  & unigrid & $0.012$ & $0.2$ and $0.1$ & Schwarzschild \\ 
 9  & unigrid & $0.049$ & $0.2$ & Schwarzschild \\ 
 10 & unigrid & $0.049$ & $0.4$ & Schwarzschild \\ 
 11 & unigrid & $0.012$ & $0.2$ & Schwarzschild \\ 
 \hline
\end{tabular}
\caption{Simulation parameters. Grid resolution $\Delta x$ refers
to the base grid resolution for the fixed mesh refinement runs.
A discussion of Schwarzschild initial data
can be found in Sec.~\ref{subsec:black_hole_initial_data},
and a discussion of ``Schwarzschild with bump'' initial data
can be found in Sec.~\ref{subsec:stability_small_pert}.
See also Sec.~\ref{sec:overview_plots}.
}
\label{table:fig_details}
\end{table}

%%%%%%%%%%%%%%%%%%%%%%%%%%%%%%%%%%%%%%%%%%%%%%%%%%%%%%%%%%%%%%%%%%%%%%%%%%%%%%
\subsection{Growth of ``hair'' from Schwarzschild initial data}
\label{subsec:black_hole_initial_data}
	For most of our simulations we begin with a ($t=const.$) slice
of the Schwarzschild black hole solution in PG coordinates;
which is (as is any spacelike slice of Schwarzschild)
an \emph{exact} solution to the initial value problem in EdGB gravity.
Specifically, at $t=0$, for $x>x_{exc}$ (the initial excision radius
as described in Sec.~\ref{subsec:evolution}), we set
\begin{equation}
\label{eqns:bh_id}
	\phi |_{t=0} = 0
	, \qquad
	Q |_{t=0} = 0
	, \qquad
	P |_{t=0} = 0
	, \qquad
	\alpha |_{t=0} = 1
	, \qquad
	\zeta |_{t=0} = \sqrt{\frac{2m}{r(x)}}
	.
\end{equation}
We then evolve this, performing a survey of outcomes varying
the EdGB coupling parameter $\lambda$ (in the GR case $\lambda=0$,
as expected, the resultant numerical solution is
static to within truncation error).

Previous studies of static scalarized black hole solutions in EdGB
gravity have found that regularity of the scalar field at the horizon places
an upper limit on the coupling parameter. For the linear coupling
case we consider, this is (see e.g.\cite{Sotiriou:2014pfa})
\begin{align}
	\frac{\lambda}{r_h^2}\leq (192)^{-1/2}
	\approx
	0.07
	,
\end{align}
where $r_h$ is the areal radius of the horizon.
For black holes much larger than this, $r_h\approx 2m$ 
(the space time is close to Schwarzschild), though approaching the extremal
limit a non-negligible amount of the spacetime mass $m$ can be
contained in the scalar field, and simply replacing
$r_h$ with $2m$ to express the above in terms
of our curvature-coupling parameter $C$ (\ref{c_coup})
gives a poor estimate of the corresponding extremal
value $C_{extr}$.
From Figure 4 of ~\cite{Sotiriou:2014pfa} we can infer a
more accurate translation :
\begin{align}
\label{eqn:extremal_curvature_coupling_SZ}
	C\leq\ C_{extr} \approx0.22
	.
\end{align}
	We find the extremal limit is not characterized by the
appearance of a naked
curvature singularity, but instead the formation of a sonic line (and
elliptic region) outside of the horizon of the black hole.
Our measured extremal limit of $C_{extr}\sim0.23$, as shown
in Sec.~\ref{subsec:internal_structure_elliptic},
is quite close to the above limit from \cite{Sotiriou:2014pfa}.

	Note that the ``extremal limit'' for a scalarized EdGB black hole
is of a different nature than the extremal limits of Kerr or
Reissner-Nordstrom black holes.
The spin or electric charge of a black hole
is set by the black hole's formation history: black holes of the same mass can
have different spins or charges depending on the initial configuration and
net charge and angular momentum of the matter that
fell in to form the black hole.
By contrast for an EdGB black hole the final scalar charge is set by the
Gauss-Bonnet coupling, and Gauss-Bonnet curvature at the horizon,
independent of its formation history
\cite{Kanti:1995vq,Sotiriou:2014pfa}.

Given that the Schwarzschild solution, of any mass, is valid initial
data in EdGB gravity, we can certainly begin with superextremal black holes
in our evolution (and again to be clear, here we use the term ``superextremal''
to refer to $C>C_{extr}$; there is no spin or charge in our numerical
solutions). As we show below however, these develop elliptic regions
outside the horizon. Moreover, our results in
\cite{Ripley:2019hxt,Ripley:2019irj}
show that trying to form a superextremal black hole from gravitational collapse
of the dilaton field (in spherical symmetry) will result in an elliptic
region appearing before
a horizon. This suggests superextremal black holes in EdGB gravity exist
in the regime of the theory governed by mixed-type equations, and their
presence or ``formation''
(however that could be interpreted in a mixed-type problem) 
would mark a breakdown of the Cauchy problem.
Also note that failure of the Cauchy problem is not {\em a priori}
connected to regions of strong curvature or black hole formation; as further
shown in \cite{Ripley:2019hxt,Ripley:2019irj}, strong coupling and 
mixed-type character can be present for arbitrarily small spacetime curvature.

\begin{figure*}
\subfloat[$C=2.5\times10^{-3}$
	\label{sfig:phi_lambda_1}]{%
	\includegraphics[width=.8\linewidth]{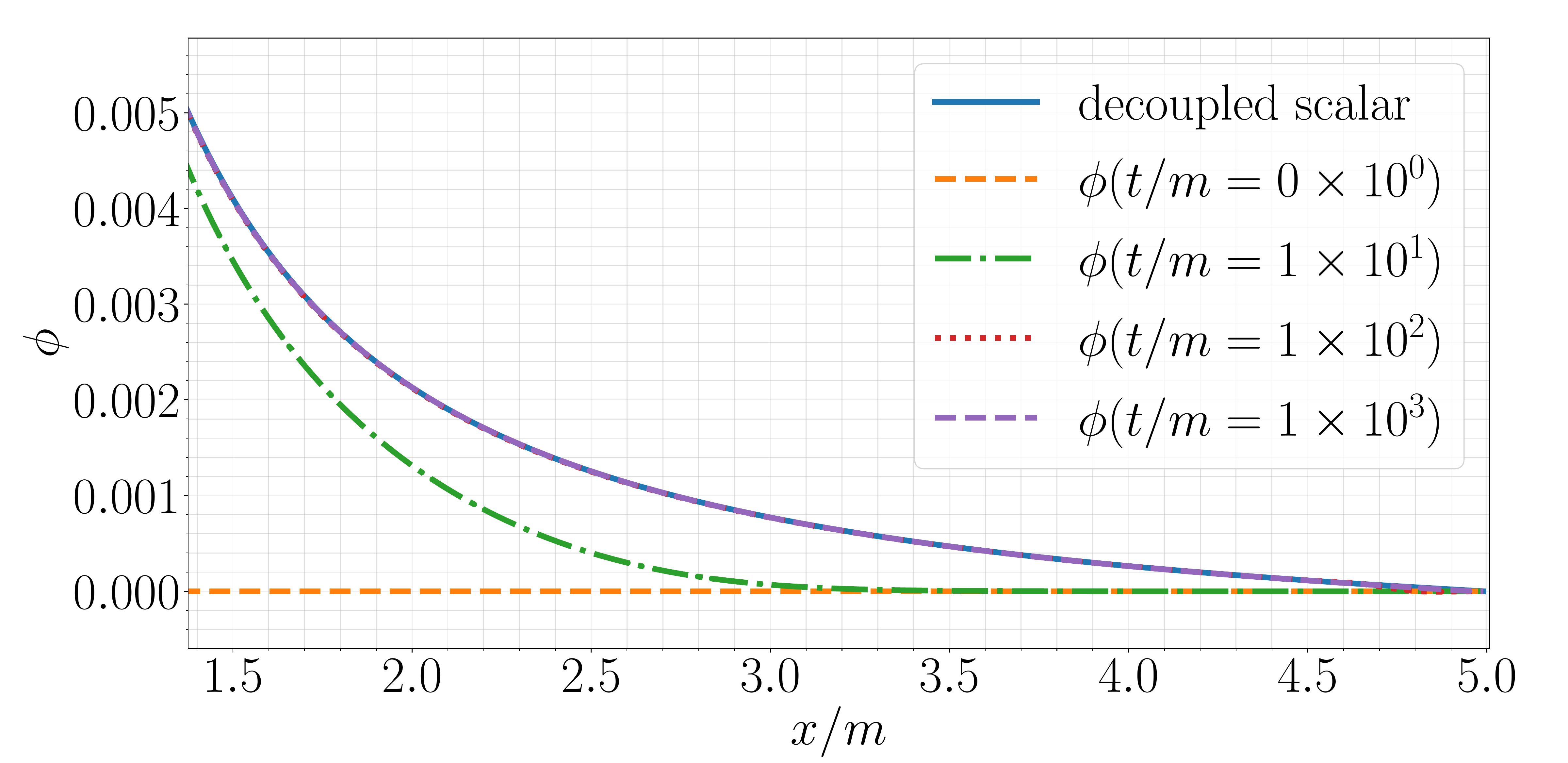}%
}\hfill
\subfloat[$C=0.16$
	\label{sfig:phi_lambda_64}]{%
	\includegraphics[width=.8\linewidth]{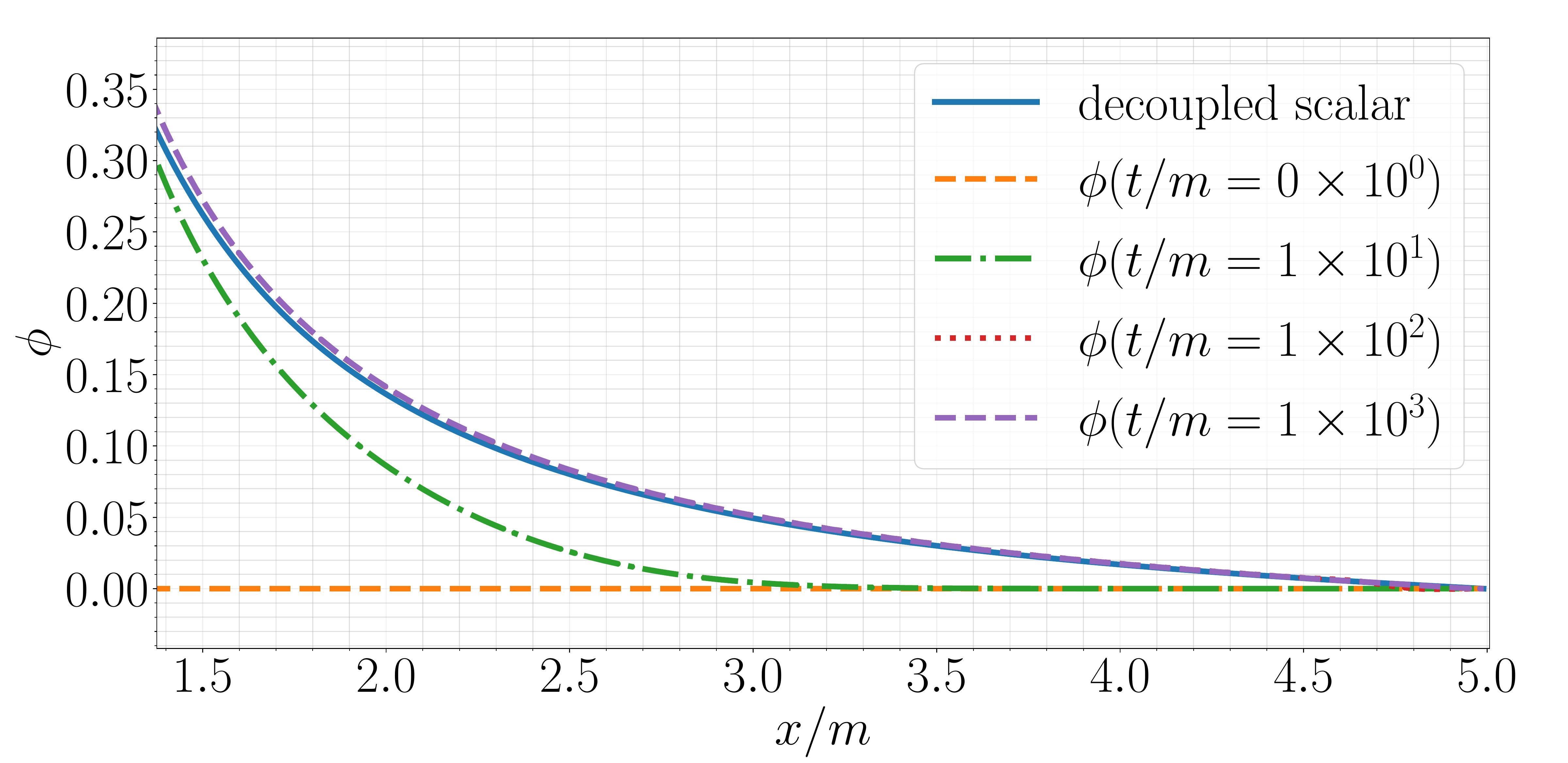}%
}
\caption{Growth of scalar ``hair'' from Schwarzschild initial data.
Shown is the scalar field profile at several times during an evolution 
for two different cases of the curvature coupling $C$ (\ref{c_coup});
the extremal limit (see the discussion in
Sec.~\ref{subsec:black_hole_initial_data}) is $C_{extr}\approx 0.23$ .
The horizon (MOTS) is located at $x_h/m\approx1.48$,
and spatial infinity is at $x/m=5$. Notice the
different range of scales on the y-axis of each figure.
Also shown for comparison is the estimate of the final profile
using  the decoupled scalar approximation (Appendix
\ref{appendix:decoupled_EdGB_scalar});
see also Fig.\ref{fig:difference_lambda_decoupled}.
For simulation parameters see Table.~\ref{table:fig_details}. 
}
\label{fig:phi_lambda_1_32}
\end{figure*}

\begin{figure*}
\includegraphics[width=.9\linewidth]{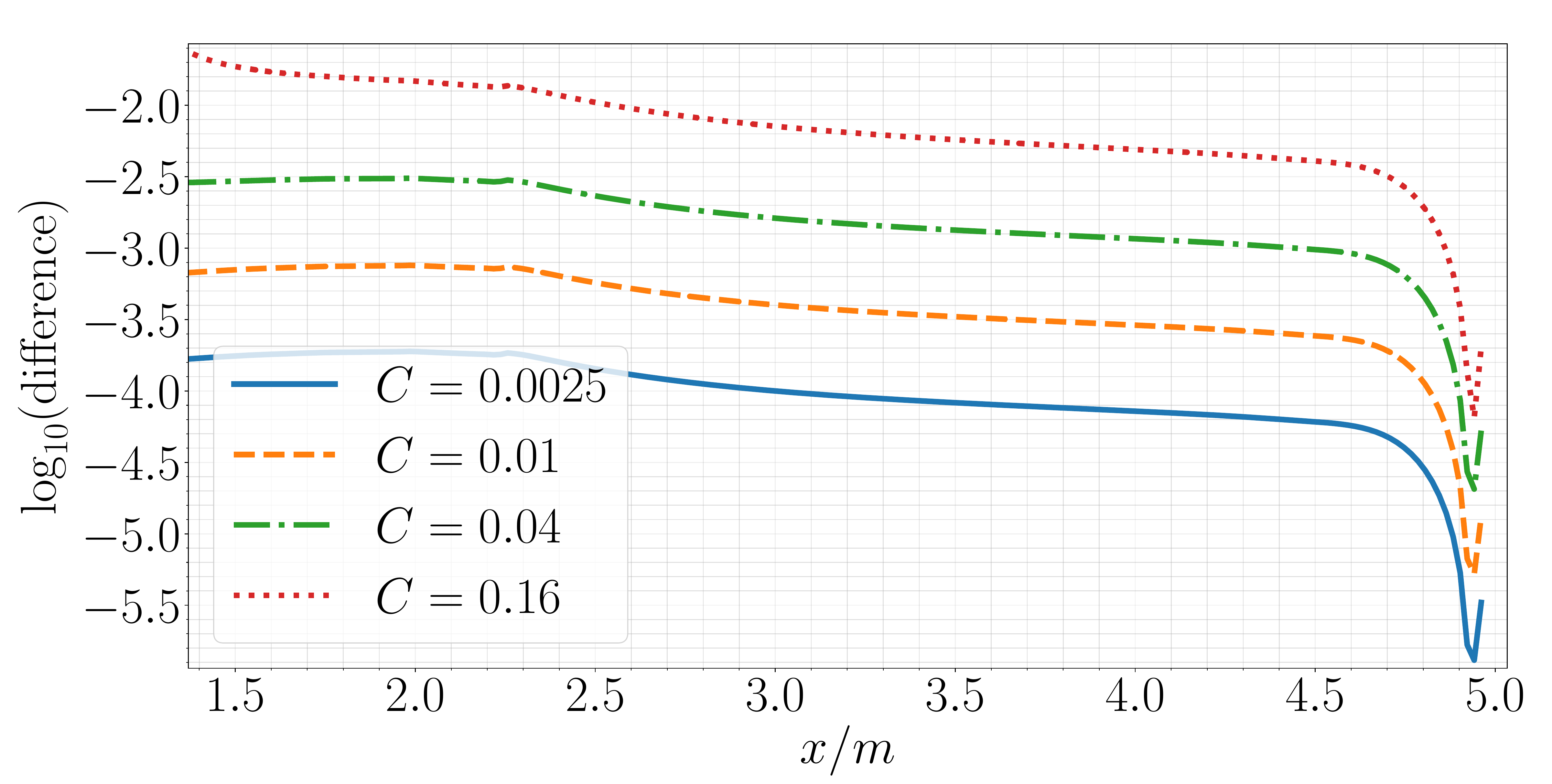}%
\caption{Difference between the late time ($t\sim2000m$)
scalar field profile obtained
from the non-linear simulations to that 
of the decoupled estimate (see Appendix \ref{appendix:decoupled_EdGB_scalar}),
for several values of the curvature coupling $C$;
see also Fig.~\ref{fig:phi_lambda_1_32}.
The black hole horizon (MOTS) is at $x/m\approx1.48$,
and spatial infinity is at $x/m=5$.
As expected, the decoupling limit approximation improves
the further $C$ is away from
the extremal limit $C_{extr}\sim 0.23$.  
For simulation parameters see Table.~\ref{table:fig_details}. 
}
\label{fig:difference_lambda_decoupled}
\end{figure*}

\begin{figure*}
\subfloat[$C=2.5\times10^{-3}$
	\label{sfig:trunc_err_phi_lambda_1}]{%
	\includegraphics[width=.9\linewidth]{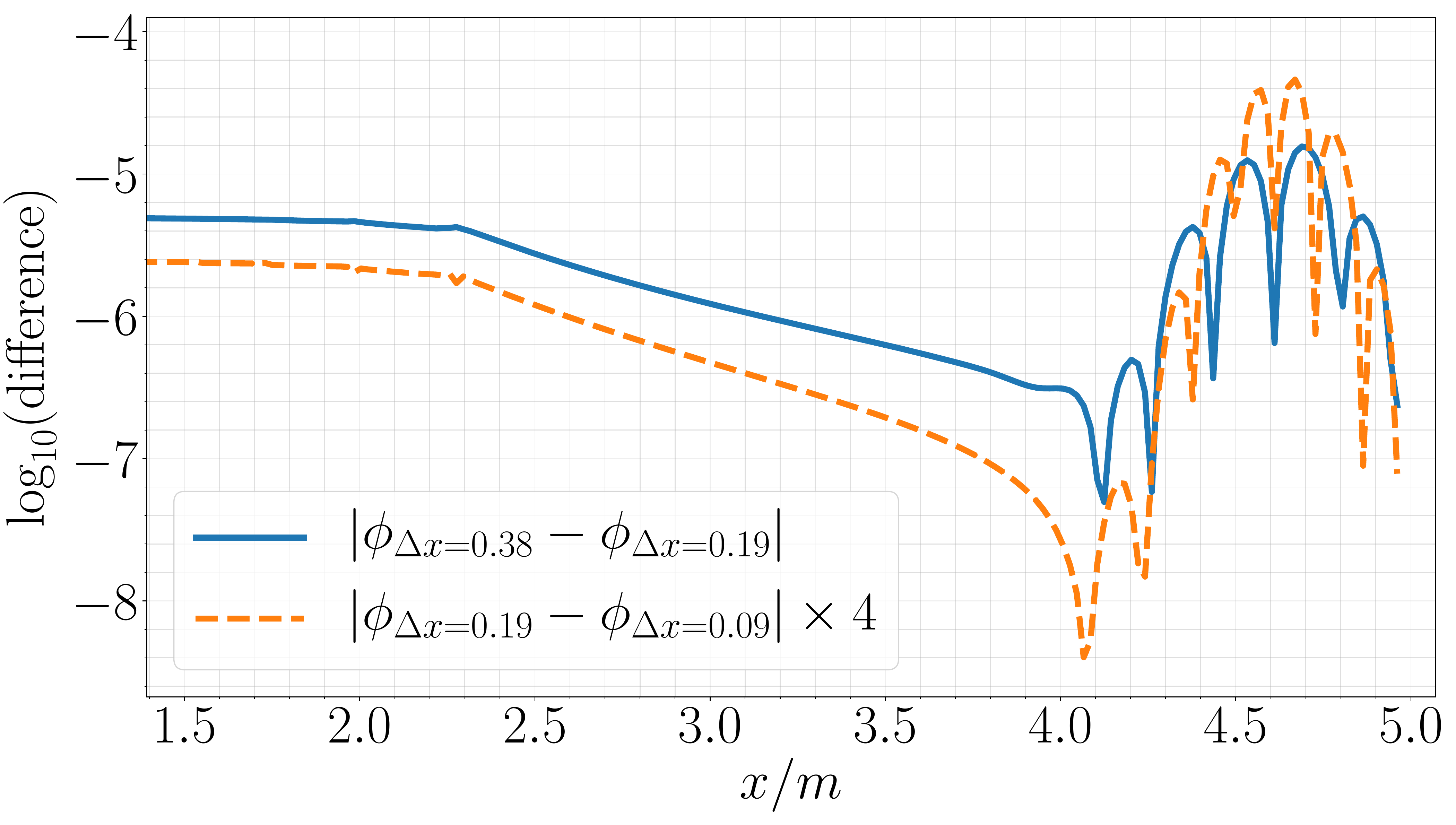}%
}\hfill
\subfloat[$C=0.16$
	\label{sfig:trunc_err_phi_lambda_64}]{%
	\includegraphics[width=.9\linewidth]{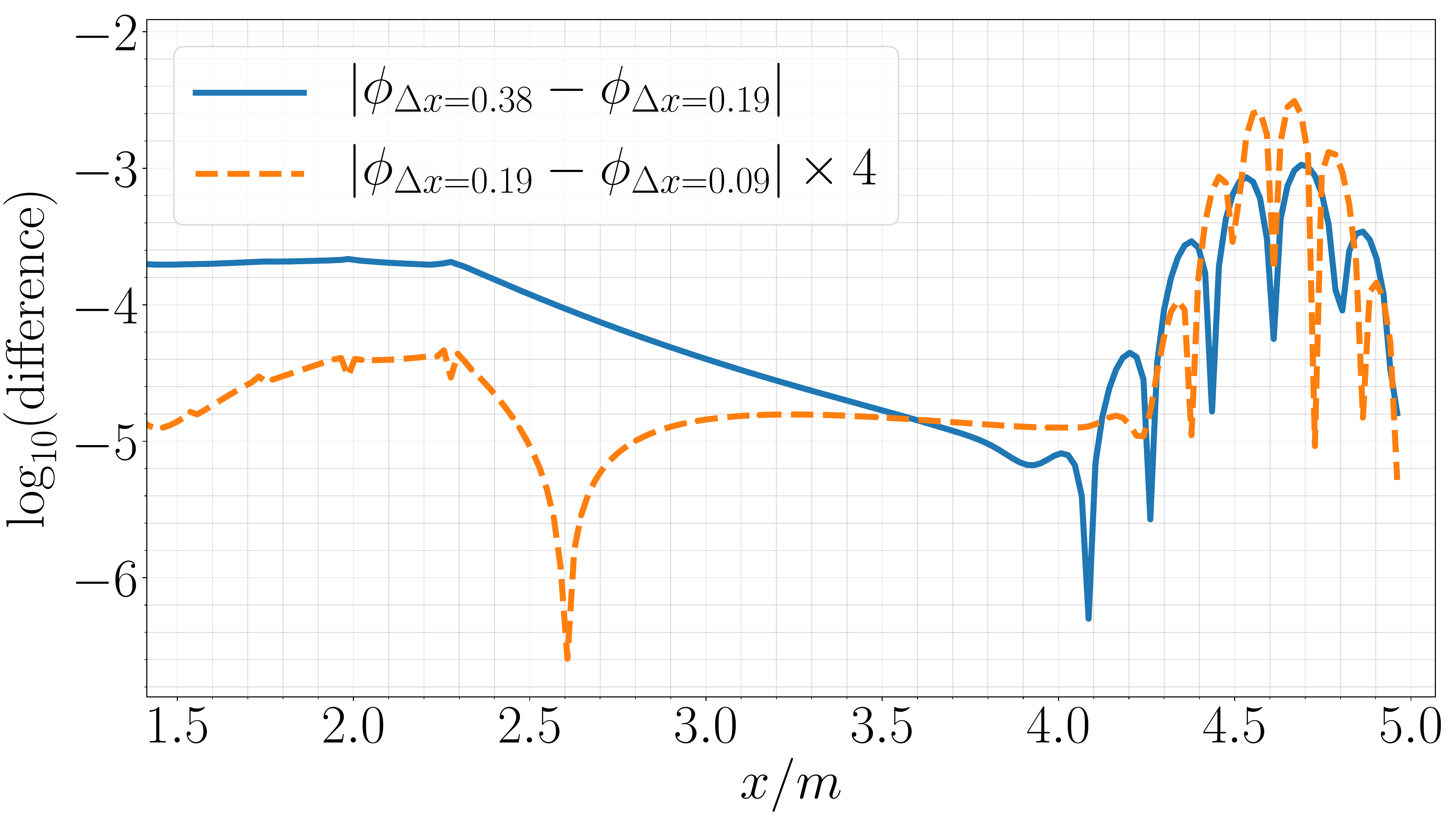}%
}
\caption{Convergence and truncation error estimate at $t\sim100m$ for
the scalar field $\phi$. Shown
are point-wise differences of the solution computed with different
resolutions; the decrease going to successively
higher resolutions is consistent with second order convergence,
and the magnitude for a given pair is an estimate of the error in the 
scalar field profile at those resolutions. 
Comparing with  
Fig.~\ref{fig:difference_lambda_decoupled}, we see
we can resolve the difference of the scalar field from the decoupled
value well above truncation error for the range of
curvature-couplings considered here.
We rescale the smaller truncation error estimate
by $4$, which is the expected
convergence rate of our code based on the order of the second order
finite difference stencils we use.
For simulation parameters see Table.~\ref{table:fig_details}. 
}
\label{fig:trunc_err_estimate_phi}
\end{figure*}

\begin{figure*}
\includegraphics[width=.9\linewidth]{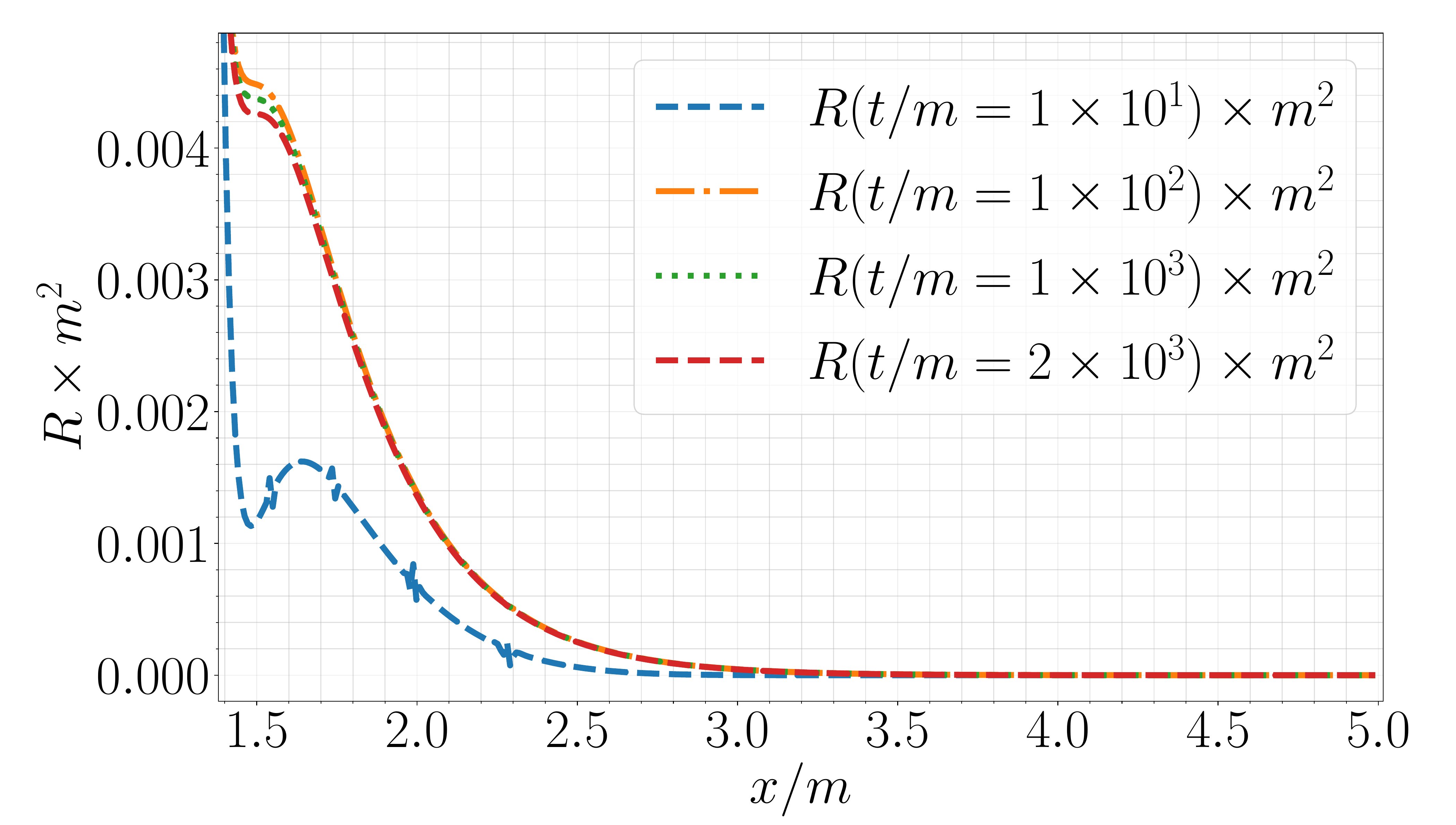}%
\caption{Evolution of the Ricci scalar for the case $C=0.16$; 
see Fig.\ref{fig:phi_lambda_1_32} for evolution of the scalar field for this
same simulation. The small ``features'' in the Ricci scalar
at the $t=10m$ time slice are located at grid refinement boundaries, and
converge away with higher base resolution
(compare with Fig.~\ref{fig:trunc_err_ricci_scalar_diff_times_lambda_64}).
For simulation parameters see Table.~\ref{table:fig_details}. 
}
\label{fig:ricci_scalar_diff_times_lambda_64}
\end{figure*}

\begin{figure*}
\includegraphics[width=.9\linewidth]{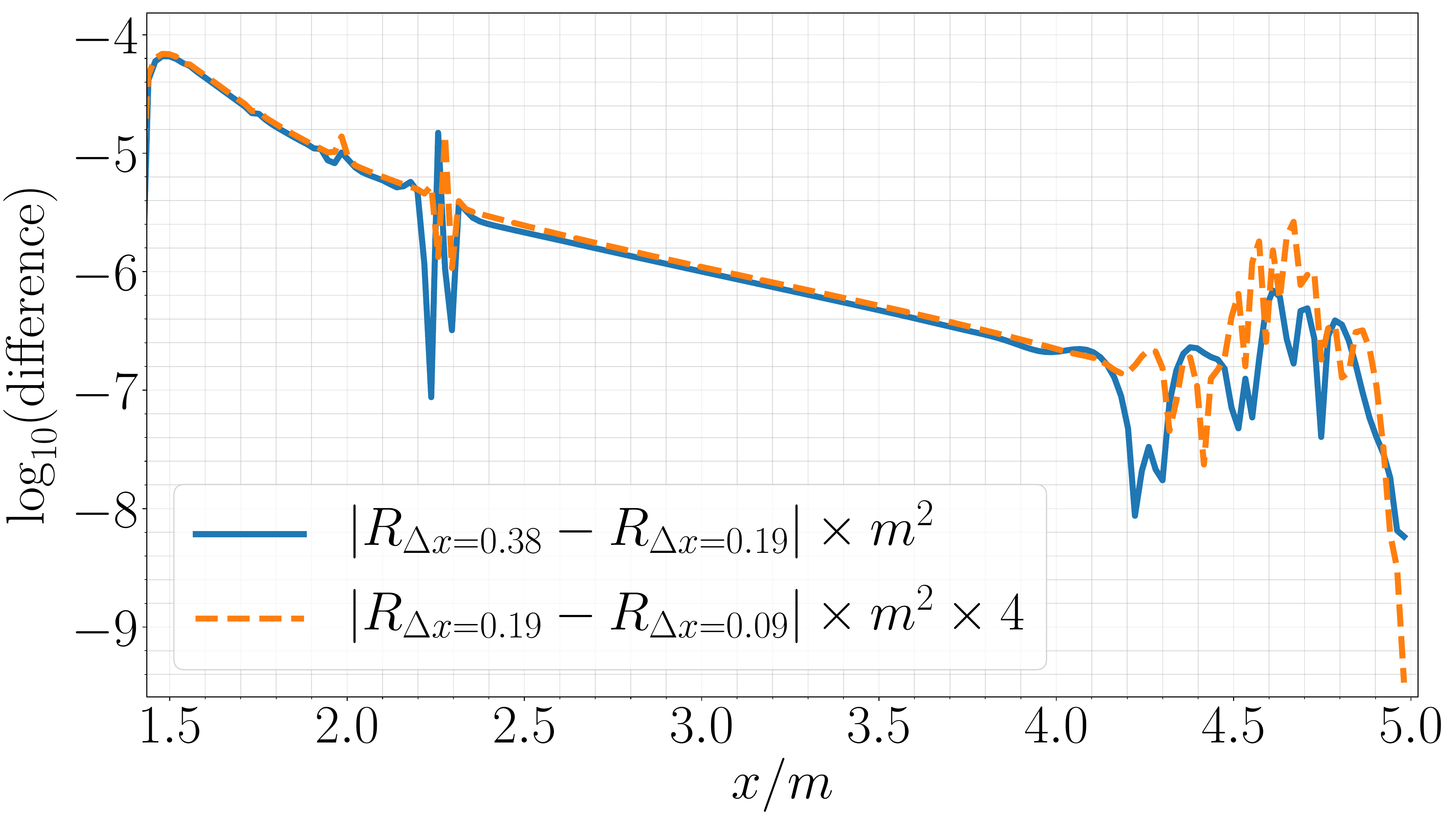}%
\caption{Convergence and truncation error estimate at $t\sim100m$ for
the Ricci scalar $R$, for the $C=0.16$ case
(see Fig.~\ref{fig:trunc_err_estimate_phi}
for a similar plot of the scalar field $\phi$ for this case,
and the caption there for a discussion of the error estimates).
The oscillations near $x/m\sim2.3$ are are at at
grid refinement boundary.
We rescale the smaller truncation error estimate by $4$, which is the expected
convergence rate of our code based on the order of the second order
finite difference stencils we use.
For simulation parameters see Table.~\ref{table:fig_details}. 
}
\label{fig:trunc_err_ricci_scalar_diff_times_lambda_64}
\end{figure*}

We first present results from evolution of Schwarzschild black hole
initial data, and curvature couplings below the extremal limit.
In all cases, if we move our excision radius sufficiently
far interior to the horizon, we find that at 
some time an elliptic region forms in the interior. However, for
these cases we can choose an excision radius closer to the horizon
so that the evolution settles to a stationary state without any
elliptic region forming in this new domain. As discussed in the introduction,
we view this as a consistent initial boundary value evolution
of EdGB gravity where the elliptic
region is ``censored'' from the exterior hyperbolic region.
	In
Figs.~\ref{fig:phi_lambda_1_32} and \ref{fig:difference_lambda_decoupled}
we show examples of scalar hair growth for these cases (with the
elliptic region excised), and their difference from the static
``decoupled'' scalar field profiles for a Schwarzschild black hole background
(see Appendix~\ref{appendix:decoupled_EdGB_scalar}),
for various curvature-couplings. These runs employed the fixed
mesh refinement algorithm
described above, with the base level grid having
$\Delta x=0.39$ resolution, Courant-Friedrichs-Lewy (CFL) number of $0.25$,
and an excision radius at fixed at $r_{exc}=0.95\times 2m$.
We find the scalar field
settles down to solutions that differ little from the static decoupled
scalar field profile, although the difference grows
as the curvature coupling approaches the extremal
limit (c.f. Fig. 3 of \cite{Sotiriou:2014pfa}). Nevertheless,
in agreement with the results of \cite{Sotiriou:2014pfa},
the difference of the full solution
from the static decoupled limit solution remains small outside
the black hole horizon.
From convergence studies we find we can 
resolve the difference of the scalar field profile from
its decoupled value well within truncation error;
see Fig.~\ref{fig:trunc_err_estimate_phi}.
For the case $C=0.16$, in Fig. \ref{fig:ricci_scalar_diff_times_lambda_64}
we show growth of Ricci curvature sourced by the scalar field,
and in Fig.~\ref{fig:trunc_err_ricci_scalar_diff_times_lambda_64}
a corresponding plot of convergence and estimated truncation error in $R$.
%%%%%%%%%%%%%%%%%%%%%%%%%%%%%%%%%%%%%%%%%%%%%%%%%
%%%%%%%%%%%%%%%%%%%%%%%%%%%%%%%%%%%%%%%%%%%%%%%%%
%%%%%%%%%%%%%%%%%%%%%%%%%%%%%%%%%%%%%%%%%%%%%%%%%
%%%%%%%%%%%%%%%%%%%%%%%%%%%%%%%%%%%%%%%%%%%%%%%%%
%%%%%%%%%%%%%%%%%%%%%%%%%%%%%%%%%%%%%%%%%%%%%%%%%
%%%%%%%%%%%%%%%%%%%%%%%%%%%%%%%%%%%%%%%%%%%%%%%%%

%%%%%%%%%%%%%%%%%%%%%%%%%%%%%%%%%%%%%%%%%%%%%%%%%%%%%%%%%%%%%%%%%%%%%%%%%%%%%%
\subsection{Perturbed Schwarzschild initial data}
\label{subsec:stability_small_pert}
\begin{figure*}
\includegraphics[width=.8\linewidth]{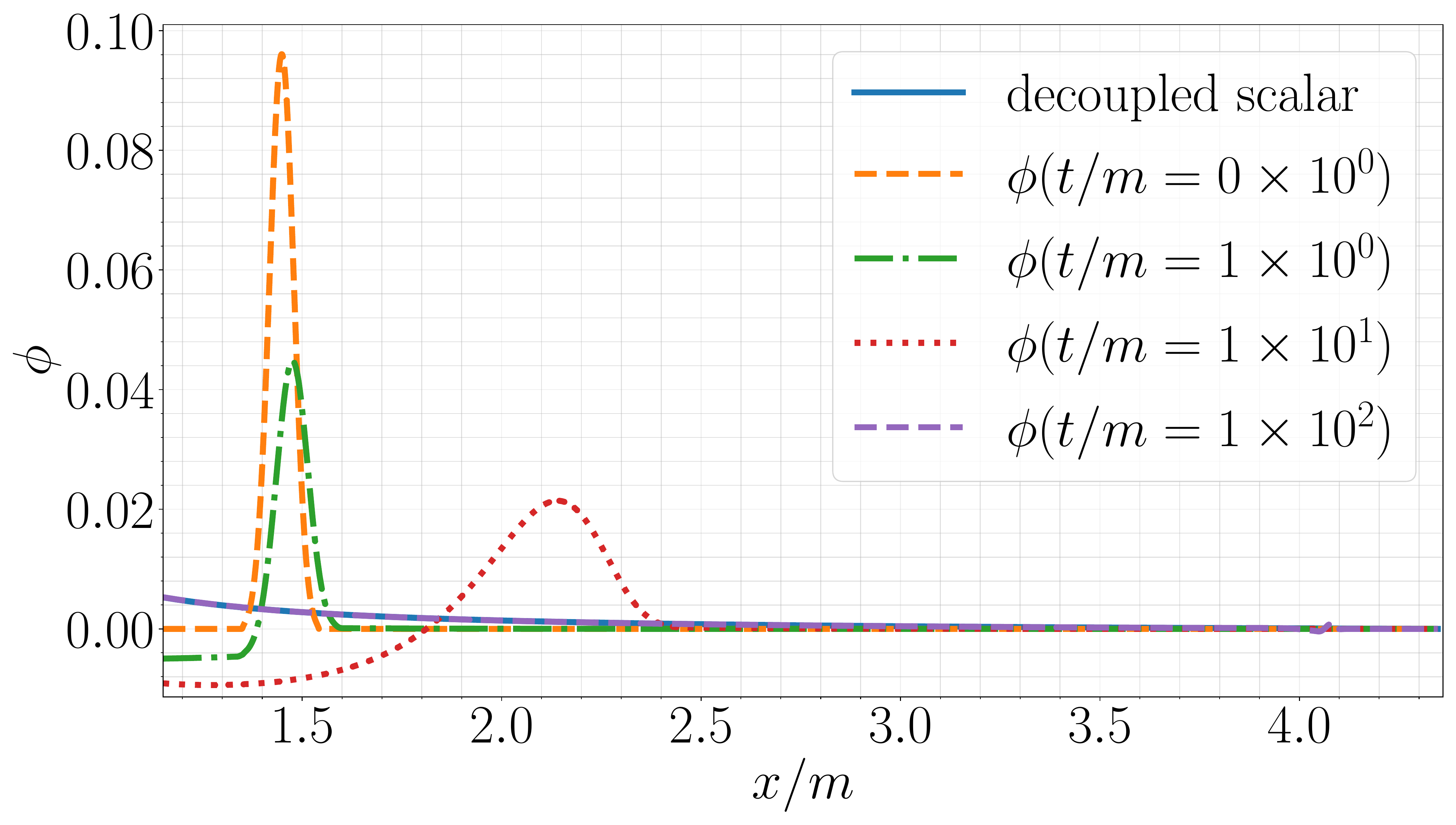}%
\caption{Schwarzschild initial data perturbed by an EdGB scalar bump
(see Eq.~\eqref{eq:bump_id}) outside the horizon. The mass of the
initial Schwarzschild black hole is $m_s=20$, while the scalar bump adds
an additional $\sim3$ in mass, giving a net mass $m=23$; thus with 
$\lambda=1$ the curvature coupling is $C=\lambda/m^2\approx2\times10^{-3}$.
Some of the scalar field falls into the black hole,
while some disperses to infinity, and at late
times the scalar field approaches the decoupled scalar
field profile.
For simulation parameters see Table.~\ref{table:fig_details}. 
} 
\label{fig:scalar_bump_example}
\end{figure*}

 Schwarzschild initial data is not generic; in particular, the scalar field
is (initially) only growing in response to the Gauss-Bonnet curvature source.
To investigate a slightly broader class of initial conditions,
here we add a small, mostly ingoing propagating perturbation to $\phi$ outside
the horizon:
\begin{subequations}
\begin{align}
\label{eq:bump_id}
	\phi(t,r)\big|_{t=0}
	= &
	\begin{cases}
	\phi_0\ \mathrm{exp}\left[-\frac{1}{(r-a)(b-r)}\right]
	\mathrm{exp}\left[-5\left(\frac{r-(a+b)/2}{a+b}\right)^2\right]
	&
	a<r<b
	\\
	0
	&
	\mathrm{otherwise}
	\end{cases}
	, \\
	Q(t,r)\big|_{t=0}
	= &
	\partial_r\phi(t,r)\big|_{t=0}
	, \\
	P(t,r)\big|_{t=0}
	= &
	0
	.
\end{align}
\end{subequations}
        This family of initial data (rescaled
``bump functions'' multiplied by a Gaussian) is smooth and
compactly supported outside the initial black hole horizon for $a>2m$.
With a fixed curvature-coupling $C$, we find that we can stably evolve 
an initial black hole plus scalar field bump if the amplitude
of the latter is sufficiently small; or equivalently if the
metric curvature measured by the Ricci scalar $R$
induced by the scalar field bump is sufficiently small.
For our initial data for $\alpha$ and $\zeta$, we set their values at the
excision surface as in Eq.~\eqref{eqns:bh_id},
and then integrate outwards in $r$.
An example of such a case is shown in Fig.~\ref{fig:scalar_bump_example}.
When the induced curvature is large, an elliptic
region forms outside the black hole horizon (and soon after that the code
crashes). As a rough estimate, we find this occurs when
$|R\times\lambda|_{\infty}\gtrsim0.1$. This result is
consistent with our earlier findings of collapse
of a scalar field pulse without any interior black hole
\cite{Ripley:2019hxt,Ripley:2019irj}.

%%%%%%%%%%%%%%%%%%%%%%%%%%%%%%%%%%%%%%%%%%%%%%%%%%%%%%%%%%%%%%%%%%%%%%%%%%%%%%

%%%%%%%%%%%%%%%%%%%%%%%%%%%%%%%%%%%%%%%%%%%%%%%%%%%%%%%%%%%%%%%%%%%%%%%%%%%%%%
\subsection{Internal structure of an EdGB black hole, and the
near extremal limit}
\label{subsec:internal_structure_elliptic}
   For small values of the curvature coupling
we can consistently excise any elliptic region that forms interior
to the horizon. We always excise the Schwarzschild curvature singularity
at $r=0$, and in none of the cases we have run (small or large curvature
coupling) do we see any signs of a curvature or scalar field singularity
forming away from $r=0$ while the equations remain hyperbolic. We can typically
evolve for some time after an elliptic region forms without excising
it before the code crashes, however this as an artifact of finite resolution,
and we can only expect convergence to the continuum limit
using our hyperbolic scheme until the first appearance of the sonic line.
In other words, to say anything rigorous about what might occur within
the elliptic region would require solving a mixed-type problem, and
we do not have the tools for doing so.

With increasing values of the curvature coupling approaching the extremal 
limit \eqref{eqn:extremal_curvature_coupling_SZ}, as the scalar field grows,
the location at which the sonic line first appears moves closer to the MOTS.
Prior to this, we excise some distance within the MOTS, though when the sonic
line appears we increase the excision radius to be at the sonic line
\footnote{The scalar and null characteristics are generally different from
each other in EdGB
gravity. For our excision strategy to be stable, we require all of the
metric and scalar characteristics to point into the excised region.
In all cases we have studied, the scalar characteristics always
are not tangent to the sonic line (the characteristics can be defined
up to the sonic line, which is also
why we classify the EdGB equations as Tricomi type here 
\cite{Ripley:2019hxt,Ripley:2019irj}).
Thus excising on the sonic line should be fine, as long as it remains
within the horizon.}. We then employ a ``high water mark'' strategy
during subsequent evolution, increasing the excision radius
to match the location of the sonic line if it grows, though do
not reduce the excision radius if the sonic line shrinks (presumed
to be happening if the characteristic discriminant on the excision boundary 
increases in magnitude away from
zero).

\begin{figure*}
\includegraphics[width=.9\linewidth]{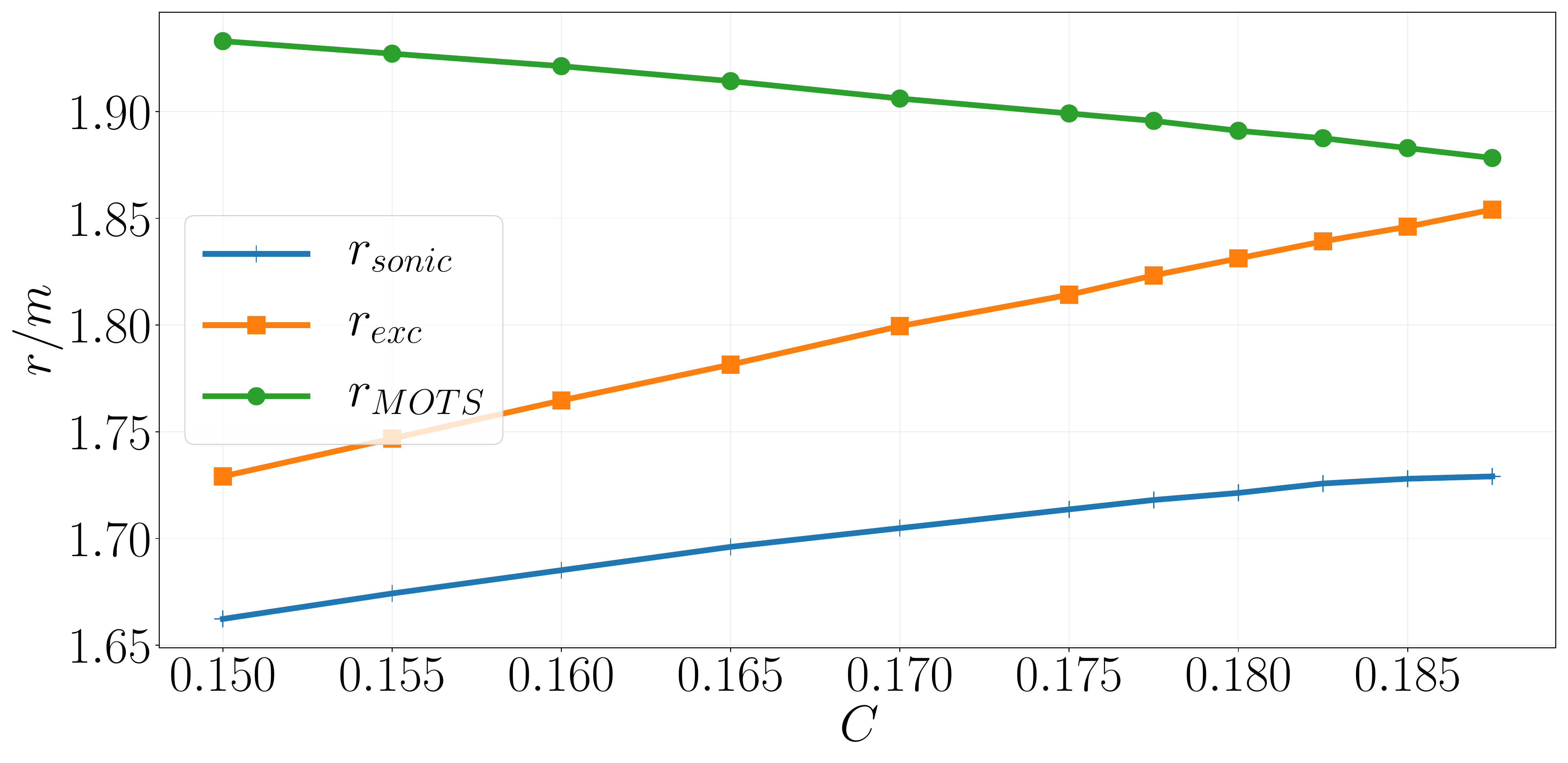}%
\caption{The location of the MOTS $r_{MOTS}$, excision radius $r_{exc}$,
and {\em estimated}
sonic line position $r_{sonic}$, as
a function of the curvature coupling $C$, measured
at $t\sim80m$ during the evolution of Schwarzschild initial data,
which is sufficiently late to give a good estimate of their
static end state values
(see Fig.~\ref{fig:size_elliptic_and_null_trapped_lambda75}).
Due to our high water mark excision strategy, the excision point represents
the largest measured 
radius the sonic line attained during evolution.
Extrapolating the curves for
the MOTS and sonic lines, we estimate the ``extremal'' coupling for
our initial data to be $C_{extr}\sim0.23$.
These results were obtained with unigrid evolution
using $\Delta x=0.012$ (corresponding to the highest
resolution curves shown in Fig.~\ref{fig:size_elliptic_and_null_trapped_lambda75}).
For $C\geq0.17$ runs the CFL number was $0.2$, while for $C<0.17$
the CFL number was $0.1$.
For other simulation parameters see Table.~\ref{table:fig_details}. 
}
\label{fig:trapped_vs_elliptic_outermost_points}
\end{figure*}

\begin{figure*}
\includegraphics[width=.9\linewidth]{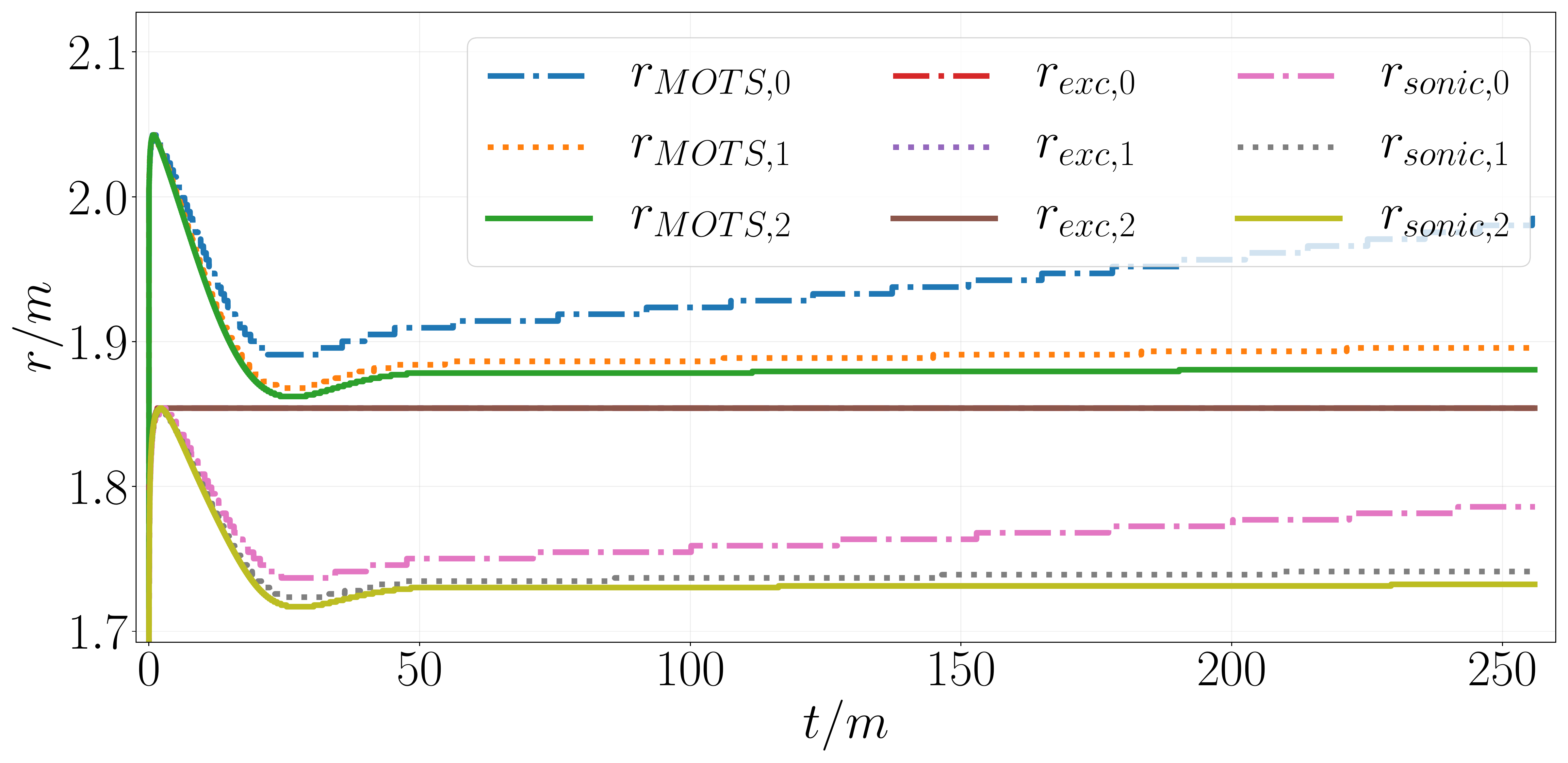}%
\caption{Evolution of the MOTS, excision point, and sonic line
as a function of time, for curvature coupling $C=0.1875$
(c.f. Fig.~\ref{fig:trapped_vs_elliptic_outermost_points}),
and from runs at three resolutions : the
labels $0,1,2$ refer to unigrid resolutions
$\Delta x=0.049$, $\Delta x=0.024$, and $\Delta x=0.012$, respectively
(a CFL factor of $0.2$ was used in all cases).
At early times as the sonic line grows, we increase the location
of the excision surface to match; after reaching a maximum
radius, the sonic line presumably starts to shrink again,
and then the curves in the figure show an estimate
of this location based on extrapolation of the characterstic
speeds (see Sec.~\ref{subsec:internal_structure_elliptic}).
The resolution study 
demonstrates that at late times we are converging 
to a static solution (in the vicinity of the horizon).
For other run parameters see Table.~\ref{table:fig_details}. 
}
\label{fig:size_elliptic_and_null_trapped_lambda75}
\end{figure*}

\begin{figure*}
\includegraphics[width=.9\linewidth]{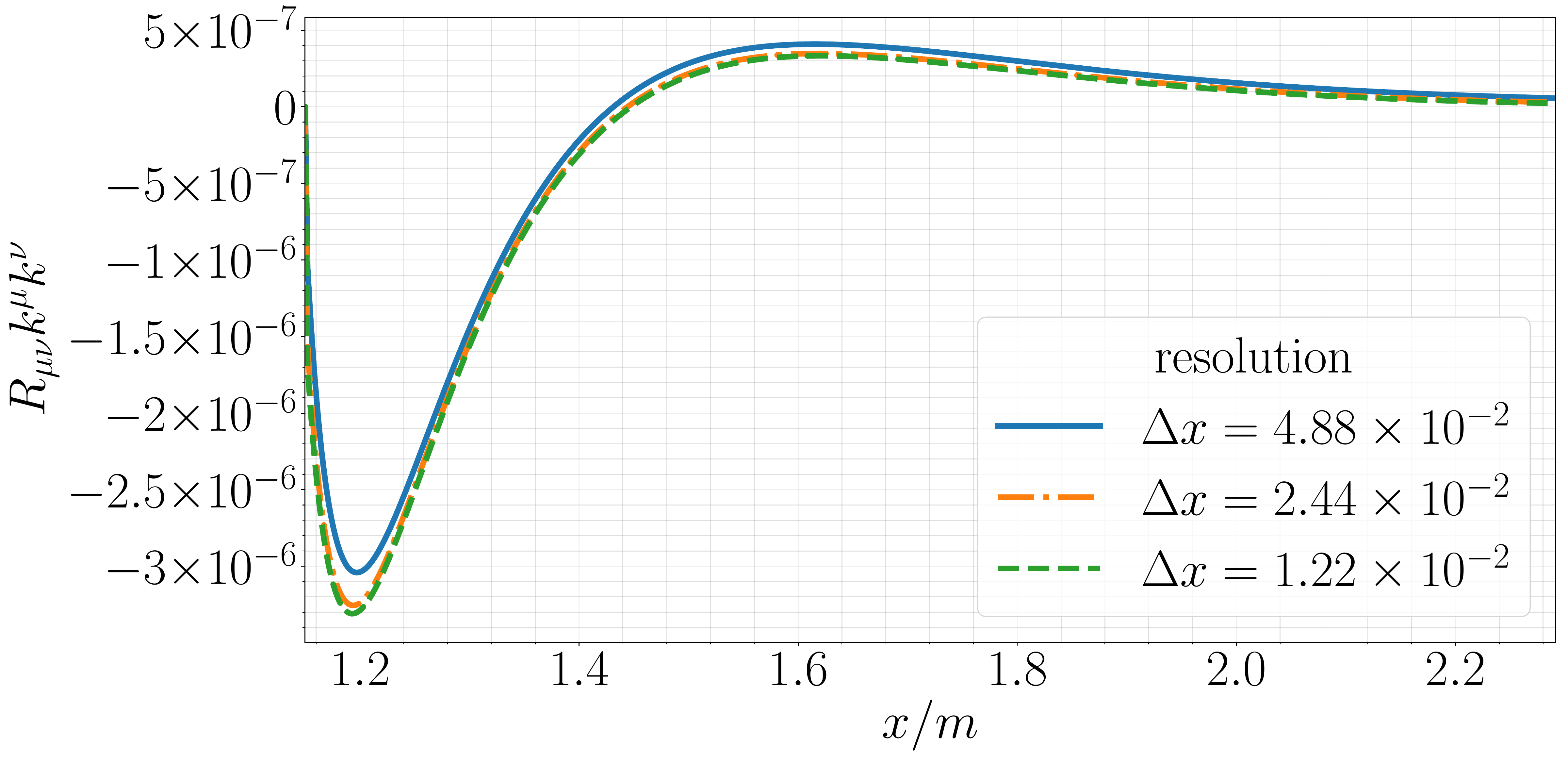}%
\caption{Ricci tensor contracted with outgoing
null vector 
$k^{\mu}=\left(1,\alpha\left(-1+\zeta\right),0,0\right)$
($R_{\mu\nu}k^{\mu}k^{\nu}$) at time $t/m=14$. 
The curvature coupling $C=0.16$.
We see that the null convergence factor
is not positive definite; where it is negative indicates
a region of NCC violation. The region of NCC violation is
localized near the black hole horizon and region
of strongest scalar field growth. This resolution study
demonstrates we can resolve the stable violation of the null convergence
condition in EdGB gravity during the formation of a scalarized
black hole solution.
For simulation parameters see Table.~\ref{table:fig_details}. 
}
\label{fig:null_convergence_condition}
\end{figure*}

For cases where the elliptic region remains censored, we typically find that 
initially the sonic line does grow, and then (presumably) shrinks within
the excision radius
as the solution settles to a stationary state. For interest,
we estimate the location of the sonic line by extrapolation,
as follows.
Recall, the equation for the radial characteristics is
(Eq.~\eqref{eq:characteristic_equation})
\begin{align}
	\mathcal{A}c^2+\mathcal{B}c+\mathcal{C}=0
	,
\end{align}
	where $\mathcal{A},\mathcal{B},\mathcal{C}$ are functions
of $\alpha,\zeta,P,Q,$ and their radial derivatives. The characteristics
thus satisfy
\begin{align}
	c_{\pm}
	=
	\frac{1}{2\mathcal{A}}\left(
	-	\mathcal{B}
	\pm	\sqrt{\mathcal{B}^2-4\mathcal{A}\mathcal{C}}
	\right)
	,
\end{align}
	and the location of
the sonic line is at the zero of the discriminant,
$\mathcal{D}\equiv\mathcal{B}^2-4\mathcal{A}\mathcal{C}$.
After excising, if $\mathcal{D}$ becomes positive definite
within the computational domain, 
we estimate the location of the sonic line as the zero
of a quadratic polynomial fitted to the function
\begin{align}
	c_+-c_-
	=
	\frac{1}{\mathcal{A}}
	\sqrt{\mathcal{B}^2-4\mathcal{A}\mathcal{C}}
	,
\end{align}
using a set of points adjacent to the excision 
boundary.
In typical cases for Schwarzschild initial data, and subextremal
curvature couplings, this estimate suggests
the true location of the final stationary sonic line lies
within $\sim94\%$ of its maximum value (the excision point);
see Fig.~\ref{fig:trapped_vs_elliptic_outermost_points} 
for a survey of the late time values of the excision radius,
MOTS location and sonic line estimate,
and Fig.~\ref{fig:size_elliptic_and_null_trapped_lambda75},
for the evolution of these quantities for one example
(including several resolutions).
In the latter figure, the shrinking of
the MOTS after some initial growth coincides
with violation of the null convergence
condition ($R_{\mu\nu}k^{\mu}k^{\nu}\geq0$ for all null
vectors $k^{\mu}$; see e.g. \cite{hawking1975large}), which is known
to occur in EdGB gravity (for more details in the spherical
collapse problem see~\cite{Ripley:2019irj}). A plot of
$R_{\mu\nu}k^{\mu}k^{\nu}$ is shown in
Fig.~\ref{fig:null_convergence_condition}.
We note that the stable violation
of the Null Convergence Condition (NCC) is thought to be a key ingredient in the construction
of singularity free cosmological and black hole solutions
(for a review, see e.g. \cite{Rubakov:2014jja}). This violation
of the NCC appears to be transient: as the
scalarized black hole settles to a stationary solution,
we find the horizon stops shrinking and the region of NCC violation
disappears. 
The slow increase in the horizon size for $t/m\gtrsim50$
is due to numerical error; we find it converges to
zero with increasing resolution.

\begin{figure*}
\includegraphics[width=.8\linewidth]{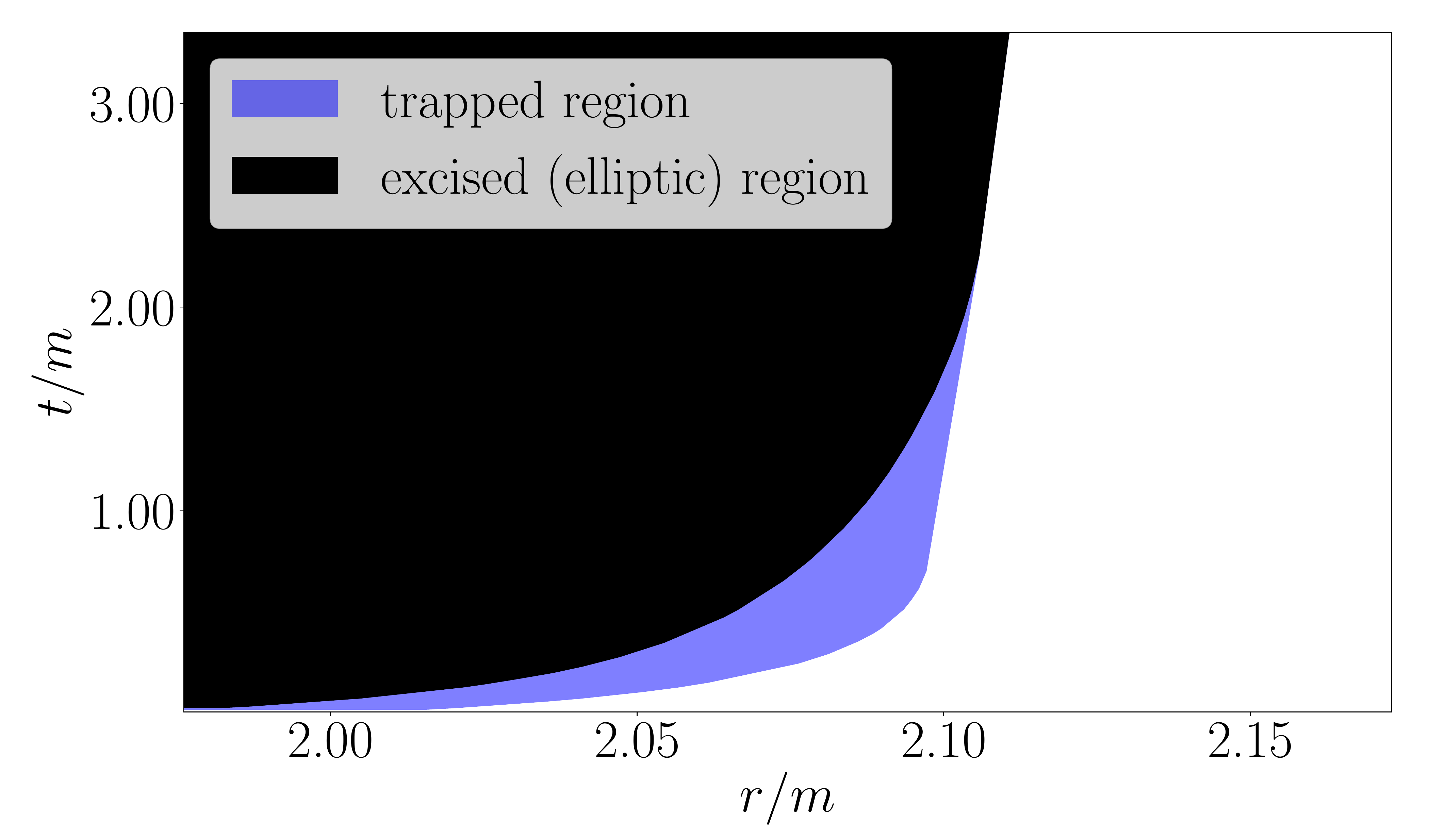}%
\caption{Example evolution of Schwarzschild initial data 
with superextremal curvature coupling: $C=0.275$.
We excise along the sonic line,
so the excised region can be thought of as the elliptic
region. The sonic line very quickly reaches and overtakes the MOTS,
and the elliptic region subsequently becomes ``naked'',
shortly after which the code crashes.
We note our code crashes if we do not excise the elliptic region,
regardless if it is interior or exterior to the black hole horizon.
For simulation parameters see Table.~\ref{table:fig_details}. 
}
\label{fig:superextremal_evolution}
\end{figure*}

For curvature couplings above $C_{extr}$, the sonic line can move
outside the MOTS, or initially appear outside it.
Linearly extrapolating the data shown
in Fig.\ref{fig:trapped_vs_elliptic_outermost_points} to the location
where the late time MOTS will cross the sonic line, we
estimate $C_{extr}\sim0.23$, close to but slightly larger than the
value $C_{extr}\sim0.22$ 
coming from seeking exactly static EdGB black hole solutions with
a non-singular $\phi$ field on the horizon\cite{Sotiriou:2014pfa} 
(though even beyond caveats with our extrapolations, we do not expect
these two methods to give identical numerical values for an extremal
coupling).
%%%%%%%%%%%%%%%%%%%%%%%%%%%%%%%%%%%%%%%%%%%%%%%%%
%%%%%%%%%%%%%%%%%%%%%%%%%%%%%%%%%%%%%%%%%%%%%%%%%
	In Fig.~\ref{fig:superextremal_evolution} we show an example evolution
of Schwarzschild initial data 
with superextremal curvature coupling. We see the sonic line quickly overtakes
the black hole horizon, leading to a ``naked'' elliptic region. 
%%%%%%%%%%%%%%%%%%%%%%%%%%%%%%%%%%%%%%%%%%%%%%%%%
%%%%%%%%%%%%%%%%%%%%%%%%%%%%%%%%%%%%%%%%%%%%%%%%%
%%%%%%%%%%%%%%%%%%%%%%%%%%%%%%%%%%%%%%%%%%%%%%%%%%%%%%%%%%%%%%%%%%%%%%%%%%%%%%
%%%%%%%%%%%%%%%%%%%%%%%%%%%%%%%%%%%%%%%%%%%%%%%%%%%%%%%%%%%%%%%%%%%%%%%%%%%%%%
\section{Discussion}
\label{sec:discussion}
	In this article we have presented numerical
evidence that in spherical symmetry,
and for sufficiently small curvature couplings (what we call subextremal),
EdGB black holes are nonlinearly stable.
For subextremal couplings even moderately close to the extremal limit,
solving the decoupled scalar equation for the
scalar field profile provides a good fit to the numerical
solution obtained in the full theory. Beginning from Schwarzschild
initial data, instead of developing a non-central
curvature singularity in the interior
as was found for static solutions~\cite{Sotiriou:2014pfa}, we find the
formation of a sonic line and elliptic region in the interior. Our treatment
of the EdGB equations as hyperbolic does not allow us to conclude
anything about possible extensions of the spacetime into the elliptic
region. For subextremal black holes, our statement about their
stability relates to the region exterior to the horizon, and
assumes that our excision strategy used to eliminate the 
interior elliptic region is self-consistent (which is supported
by the stability/convergence of the corresponding numerical evolutions).
For superextremal cases, the sonic line forms or evolves to be outside
the horizon, meaning we cannot excise it, and we would need to treat the
exterior equations as mixed-type to obtain sensible solutions (or said another
way, then the exterior evolution ceases to satisfy a well-posed
Cauchy initial value problem). The particular value of the curvature
coupling we find for the extremal limit is similar to, but slightly
different from that given for static EdGB black holes 
solutions~\cite{Kanti:1995vq,Sotiriou:2014pfa}; this is not particularly
surprising given we are dynamically forming them from Schwarzschild
initial data.

	There are various ways in which this work could be extended. One
is to explore a wider class of initial conditions; for example,
collapse to a black hole from a regular matter source, 
whether the pure EdGB scalar field as in~\cite{Ripley:2019irj}, 
or coupled to another source of matter driving most of the
collapse (e.g. extending the study of~\cite{Benkel:2016rlz}, which only 
considered the decoupled EdGB field on top of Oppenheimer-Snyder style collapse,
to the full EdGB equations). Recent
work suggests that
whether or not scalarized black holes form in the theory depends on the
functional form of $f(\phi)$ (see Eq.~\eqref{eq:EdGBAction})
\cite{PhysRevD.99.064011,PhysRevD.99.044017}; with the methods
presented in this paper one could explore
these questions with numerical solutions to the full
theory in spherical symmetry. Another future direction is to study
critical collapse in EdGB gravity using adaptive mesh refinement. 

	Finally, this work could be extended by considering numerical
solutions of EdGB gravity in axisymmetry, or without any symmetry restrictions.
This would couple in propagating metric degrees of freedom, and hence introduce
a qualitatively different aspect of the theory not available in spherical
symmetry. If, similar to the conclusions found here and
in \cite{Ripley:2019hxt,Ripley:2019irj}, there exist
subsets of initial data that offer well-posed hyperbolic evolution,
then EdGB gravity may still be viable as an interesting modified
gravity theory to confront with gravitational wave binary merger
data.  On the other hand, if 
the linear analysis in~\cite{Papallo:2017qvl,Papallo:2017ddx} 
that EdGB gravity is generically ill-posed in a particular gauge
applies to all gauges, then the well-posed
cases we have found could be an artifact of spherical symmetry,
and including any gravitational wave degrees of freedom
would render the theory ill-posed.

%%%%%%%%%%%%%%%%%%%%%%%%%%%%%%%%%%%%%%%%%%%%%%%%%%%%%%%%%%%%%%%%%%%%%%%%%%%%%%
\begin{acknowledgments}
	We thank Leo Stein for several encouraging
discussions while this project was underway, and the anonymous referee
for their helpful comments on an earlier draft of this paper.
Computational resources were provided courtesy of the Feynman cluster at
Princeton University. F.P. acknowledges support from NSF
grant PHY-1912171, the Simons Foundation, and the
Canadian Institute For Advanced Research (CIFAR).
\end{acknowledgments}

\appendix
%%%%%%%%%%%%%%%%%%%%%%%%%%%%%%%%%%%%%%%%%%%%%%%%%%%%%%%%%%%%%%%%%%%%%%%%%%%%%%
\section{Convergence of an independent residual}
\label{appendix:convergence}
\begin{figure*}
\includegraphics[width=.8\linewidth]{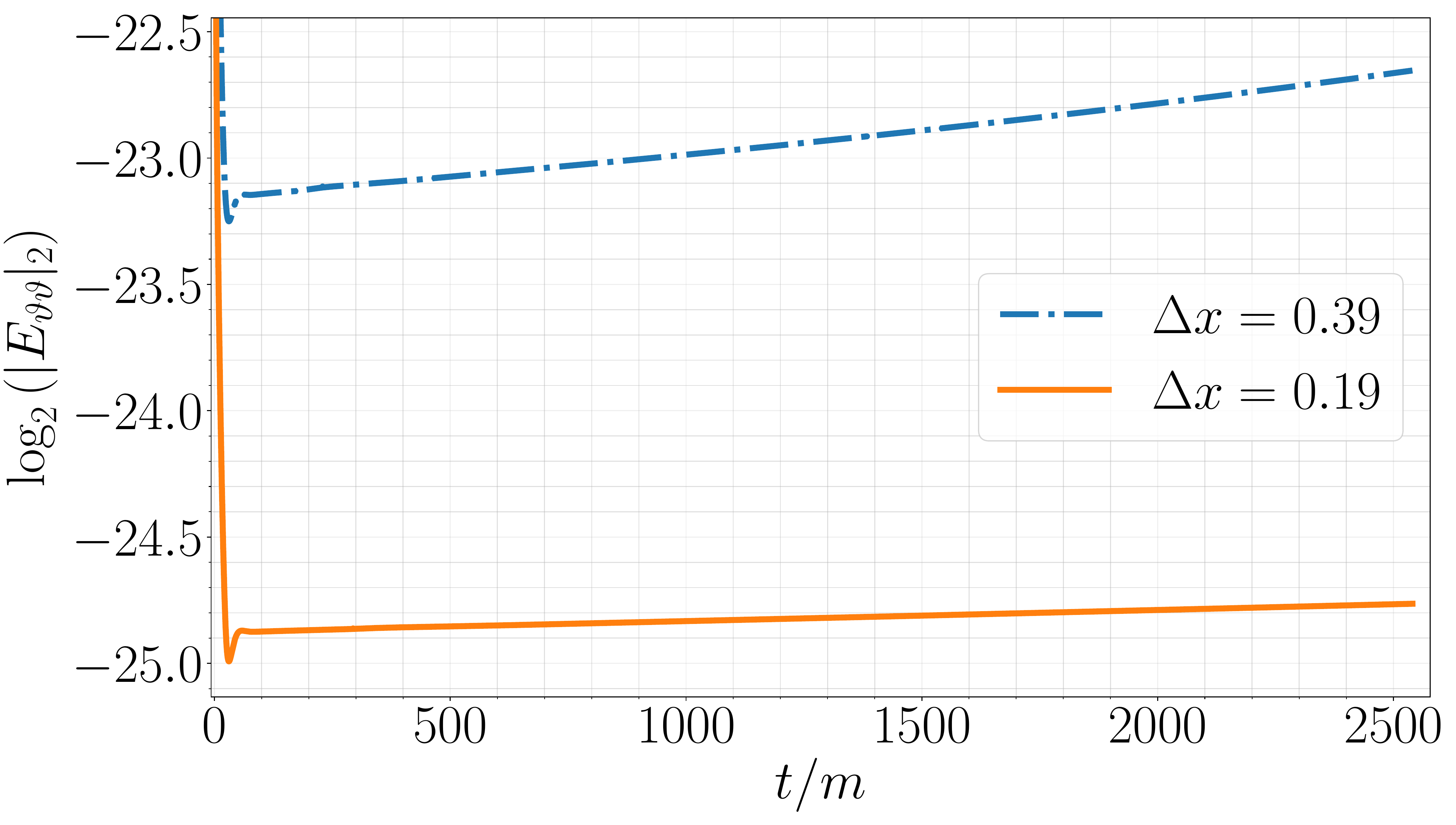}%
\caption{Second order convergence to zero of the $\vartheta\vartheta$ component
of the equation of motion (\ref{eq:tensor_eom}) from a representative run:
Schwarzschild black hole initial data, and curvature coupling $C=0.16$. 
Shown is $\mathrm{log}_2\left(|E_{\vartheta\vartheta}|_2\right)$
versus time computed
on the base level of a fixed mesh refinement run (other levels also show
second order convergence to zero). The $\mathrm{log}_2$ is chosen so the second
order convergence is more apparent.
For simulation parameters see Table.~\ref{table:fig_details}. 
}
\label{fig:convergence_ThTh}
\end{figure*}

	In Fig.~\eqref{fig:convergence_ThTh} we present the two norm of the
$E_{\vartheta\vartheta}$ component of the equation of motion for
a representative case,
excising the elliptic region, and
evolved with fixed mesh refinement.
We see second order convergence to zero over the entire run-time
$t\approx2\times10^3m$ of the simulation.
The plot only shows the norm computed on the coarsest level,
although we observe second order convergence over all levels of fixed mesh 
refinement (four in addition to the base level).
As an example of how this translates to solution error, for the
highest resolution case shown in Fig.\ref{fig:convergence_ThTh},
after the early time transient behavior and the solution
has settled to be nearly static at the horizon
(see Fig.\ref{fig:size_elliptic_and_null_trapped_lambda75}), we see 
a net drift in the mass of the black hole
of $\delta m/m\sim0.4\%$ over the remainder of the simulation.
Other curvature couplings give similar results. 
If we do not excise the elliptic region we begin to loose convergence there, and
eventually the code crashes, as expected. 

%%%%%%%%%%%%%%%%%%%%%%%%%%%%%%%%%%%%%%%%%%%%%%%%%%%%%%%%%%%%%%%%%%%%%%%%%%%%%%
\section{Decoupled EdGB scalarized black hole solutions}
\label{appendix:decoupled_EdGB_scalar}
	For reference, we present the decoupled scalar solutions for
EdGB gravity in PG coordinates. The decoupling limit for EdGB gravity
is the solution of the scalar wave equation
\begin{equation}
\label{eq:decoupled_scalar_equation}
	\Box\phi+\lambda\mathcal{R}_{GB} = 0
	,
\end{equation}
	about a GR background. Our background is the Schwarzschild spacetime:
\begin{align}
	\alpha
	=
	1
	,\qquad
	\zeta
	=
	\sqrt{\frac{2m}{r}}
	.
\end{align}
	We assume a static solution, $\phi(r)$, for which
Eq.~\eqref{eq:decoupled_scalar_equation} becomes
\begin{equation}
	\frac{1}{r^2}\frac{d}{dr}\left(
		r^2\left(1-\frac{2m}{r}\right)\frac{d\phi}{dr}
	\right)
+	\frac{48\lambda m^2}{r^6}
	=
	0
	.
\end{equation} 
	The solution to this equation is 
\begin{equation}
	\phi(r)
	=
	\frac{2\lambda}{m}\frac{1}{r}
+	2\lambda\frac{1}{r^2}
+	\frac{8m\lambda}{3}\frac{1}{r^3}
+	\frac{c_1m+2\lambda}{2m^2}\mathrm{log}\left(1-\frac{2m}{r}\right)
+	c_2
	,
\end{equation}
	where $c_1$ and $c_2$ are integration constants. Regularity at the
black hole horizon sets $c_1=-2\lambda/m$, and requiring $\phi(r=\infty)=0$
sets $c_2=0$. We have 
\begin{align}
\label{eq:static_decoupled_phi}
	\phi(r)
	=
	\frac{2\lambda}{m^2}\left(
		\frac{m}{r}
	+	\frac{m^2}{r^2}
	+	\frac{4}{3}\frac{m^3}{r^3}
	\right)	
	,	
\end{align}
	which is the solution we compare our numerical results
against in this article. Note from \cite{Benkel:2016kcq} that the dynamical
solution to Eq.~\eqref{eq:decoupled_scalar_equation} has been shown to
asymptotically settle to the static solution
\eqref{eq:static_decoupled_phi} for Schwarzschild backgrounds. 
%%%%%%%%%%%%%%%%%%%%%%%%%%%%%%%%%%%%%%%%%%%%%%%%%%%%%%%%%%%%%%%%%%%%%%%%%%%%%%
\section{Form of functions
$\mathcal{A}_{(P)}$$\mathcal{F}_{(P)}$,
$\mathcal{A}_{(\zeta)}$, and $\mathcal{F}_{(\zeta)}$}
\label{appendix:form_of_long_terms}
	Here we provide the (lengthy) expressions for the functions
$\mathcal{A}_{(P)}$ and $\mathcal{F}_{(P)}$,
(see Eq.~\eqref{eq:evolution_P})$,
\mathcal{A}_{(\zeta)}$, and $\mathcal{F}_{(\zeta)}$
(see Eq.~\eqref{eq:free_evolution_zeta}),
which we produced using Mathematica \cite{Mathematica}.
While we work only for $f(\phi)=\phi$ in this article, we show
the complete expressions for reference. 
\begin{subequations}
\begin{align}
	&\mathcal{A}_{(P)}
	\equiv
	1
+	\left(
		-\frac{16 \lambda  Q}{r}-\frac{16 \lambda  P \zeta }{r}
	\right) f^{\prime}
+	\left(
		\frac{64 \lambda^2 Q^2}{r^2}
	+	\frac{128 \lambda^2 P Q \zeta }{r^2}
	+	\frac{64 \lambda^2 P^2 \zeta^2}{r^2}
	-	\frac{32 \lambda^2 \zeta^4}{r^4}
	\right) \left(f^{\prime}\right)^2
	\nonumber\\&
+	\frac{256 \lambda^3 Q^2 \zeta^4
	\left(f^{\prime}\right)^2 f^{\prime\prime}}{r^4}
+	\frac{256 \lambda^3 \zeta^4 \left(f^{\prime}\right)^3 \partial_r Q}{r^4}
+	\left(
		\frac{
			128 \lambda^2 \zeta^4 \left(f^{\prime}\right)^2
		}{r^3 \alpha }
	+	\left(
		-	\frac{1024 \lambda^3 Q \zeta^4}{r^4 \alpha }
		-	\frac{768 \lambda^3 P \zeta^5}{r^4 \alpha }
		\right)\left(f^{\prime}\right)^3
	\right) \partial_r\alpha
	\nonumber\\&
+	\left(
		\frac{128 \lambda^2 \zeta^3 \left(f^{\prime}\right)^2}{r^3}
	+	\left(
		-	\frac{1024 \lambda^3 Q \zeta^3}{r^4}
		-	\frac{768 \lambda^3 P \zeta^4}{r^4}
		\right)\left(f^{\prime}\right)^3
	\right) \partial_r\zeta
	,\\
	&\mathcal{F}_{(P)}
	\equiv
	-\frac{2 Q \alpha }{r}
-	\frac{2 P \alpha  \zeta }{r}
+	\left(
		\frac{32 \lambda  Q^2 \alpha }{r^2}
	+	\frac{60 \lambda  P Q \alpha  \zeta }{r^2}
	+	\frac{34 \lambda  P^2 \alpha  \zeta^2}{r^2}
	-	\frac{6 \lambda  Q^2 \alpha  \zeta^2}{r^2}
	-	\frac{4 \lambda  \alpha  \zeta^4}{r^4}\right) f^{\prime}
	\nonumber\\&
+	\left(
	-	\frac{128 \lambda^2 Q^3 \alpha }{r^3}
	-	\frac{384 \lambda^2 P Q^2 \alpha  \zeta }{r^3}
	-	\frac{416 \lambda^2 P^2 Q \alpha  \zeta^2}{r^3}
	+	\frac{32 \lambda^2 Q^3 \alpha  \zeta^2}{r^3}
	-	\frac{160 \lambda^2 P^3 \alpha  \zeta^3}{r^3}
	+	\frac{32 \lambda^2 P Q^2 \alpha  \zeta^3}{r^3}
	\right) \left(f^{\prime}\right)^2
	\nonumber\\&
+	\left(
		\frac{32 \lambda^2 P Q^3 \alpha  \zeta }{r^2}
	+	\frac{16 \lambda^2 P^2 Q^2 \alpha  \zeta^2}{r^2}
	+	\frac{16 \lambda^2 Q^4 \alpha  \zeta^2}{r^2}
	-	\frac{32 \lambda^2 P Q \alpha  \zeta^3}{r^4}
	-	\frac{32 \lambda^2 P^2 \alpha  \zeta^4}{r^4}
	+	\frac{32 \lambda^2 Q^2 \alpha  \zeta^4}{r^4}
	\right) f^{\prime} f^{\prime\prime}
	\nonumber\\&
+	\left(
		\frac{256 \lambda^3 P Q^3 \alpha  \zeta^3}{r^4}
	+	\frac{256 \lambda^3 P^2 Q^2 \alpha  \zeta^4}{r^4}
	\right) f^{\prime} \left(f^{\prime\prime}\right)^2
	\nonumber\\&
+	\Bigg(
	-	\alpha 
	+	\left(
			\frac{16 \lambda  Q \alpha }{r}
		+	\frac{16 \lambda  P \alpha  \zeta }{r}
		\right) f^{\prime}
	\nonumber\\&
	+	\left(
		-	\frac{64 \lambda^2 Q^2 \alpha }{r^2}
		-	\frac{96 \lambda^2 P Q \alpha  \zeta }{r^2}
		-	\frac{48 \lambda^2 P^2 \alpha  \zeta^2}{r^2}
		+	\frac{16 \lambda^2 Q^2 \alpha  \zeta^2}{r^2}
		+	\frac{32 \lambda^2 \alpha  \zeta^4}{r^4}
		\right) \left(f^{\prime}\right)^2
	\nonumber\\&
	+	\left(
			\frac{256 \lambda^3 P Q \alpha  \zeta^3}{r^4}
		+	\frac{256 \lambda^3 P^2 \alpha  \zeta^4}{r^4}
		\right) \left(f^{\prime}\right)^2 f^{\prime\prime}
	\Bigg) \partial_r Q
	\nonumber\\&
+	\Bigg(
	-	\alpha  \zeta 
	+	\left(
			\frac{16 \lambda  Q \alpha  \zeta }{r}
		+	\frac{16 \lambda  P \alpha  \zeta^2}{r}
		\right) f^{\prime}
	\nonumber\\&
	+	\left(
		-	\frac{64 \lambda^2 Q^2 \alpha  \zeta }{r^2}
		-	\frac{128 \lambda^2 P Q \alpha  \zeta^2}{r^2}
		-	\frac{32 \lambda^2 \alpha  \zeta^3}{r^4}
		-	\frac{64 \lambda^2 P^2 \alpha  \zeta^3}{r^2}
		+	\frac{32 \lambda^2 \alpha  \zeta^5}{r^4}
		\right) \left(f^{\prime}\right)^2
	\nonumber\\&
	+	\left(
			\frac{256 \lambda^3 Q^2 \alpha  \zeta^3}{r^4}
		-	\frac{256 \lambda^3 Q^2 \alpha  \zeta^5}{r^4}
		\right) \left(f^{\prime}\right)^2 f^{\prime\prime}
	+	\left(
			\frac{256 \lambda^3 \alpha  \zeta^3}{r^4}
		-	\frac{256 \lambda^3 \alpha  \zeta^5}{r^4}
		\right) \left(f^{\prime}\right)^3 \partial_r Q
	\Bigg)\partial_r P 
	\nonumber\\&
+	\Bigg(
		\left(
			\frac{64 \lambda^2 P \zeta^3}{r^3 \alpha }
		+	\frac{64 \lambda^2 Q \zeta^4}{r^3 \alpha }
		\right) \left(f^{\prime}\right)^2
	\nonumber\\&
	+	\left(
		-	\frac{512 \lambda^3 P Q \zeta^3}{r^4 \alpha }
		-	\frac{512 \lambda^3 P^2 \zeta^4}{r^4 \alpha }
		-	\frac{512 \lambda^3 Q^2 \zeta^4}{r^4 \alpha }
		-	\frac{512 \lambda^3 P Q \zeta^5}{r^4 \alpha }
		\right) \left(f^{\prime}\right)^3
	\Bigg) \left(\partial_r\alpha\right)^2
	\nonumber\\&
+	\Bigg(
	-	P \alpha 
	+	\left(
			\frac{24 \lambda  P Q \alpha }{r}
		+	\frac{20 \lambda  P^2 \alpha  \zeta }{r}
		+	\frac{4 \lambda  Q^2 \alpha  \zeta }{r}
		+	\frac{16 \lambda  \alpha  \zeta^3}{r^3}
		\right) f^{\prime}
	\nonumber\\&
	+	\left(
		-	\frac{128 \lambda^2 P Q^2 \alpha }{r^2}
		-	\frac{192 \lambda^2 P^2 Q \alpha  \zeta }{r^2}
		-	\frac{32 \lambda^2 Q^3 \alpha  \zeta }{r^2}
		-	\frac{80 \lambda^2 P^3 \alpha  \zeta^2}{r^2}
		-	\frac{16 \lambda^2 P Q^2 \alpha  \zeta^2}{r^2}
		-	\frac{160 \lambda^2 Q \alpha  \zeta^3}{r^4}
		-	\frac{96 \lambda^2 P \alpha  \zeta^4}{r^4}
		\right) \left(f^{\prime}\right)^2
	\nonumber\\&
	+	\left(
			\left(
				\frac{64 \lambda^2 P Q \alpha  \zeta^2}{r^3}
			+	\frac{128 \lambda^2 P^2 \alpha  \zeta^3}{r^3}
			\right) f^{\prime}
		+	\left(
			-	\frac{512 \lambda^3 P Q^2 \alpha  \zeta^2}{r^4}
			-	\frac{1280 \lambda^3 P^2 Q \alpha  \zeta^3}{r^4}
			+	\frac{256 \lambda^3 Q^3 \alpha  \zeta^3}{r^4}
			-	\frac{768 \lambda^3 P^3 \alpha  \zeta^4}{r^4}
			\right) \left(f^{\prime}\right)^2
		\right)f^{\prime\prime}
	\nonumber\\&
	+	\left(
			\left(
				\frac{64 \lambda^2 \alpha  \zeta^2}{r^3}
			-	\frac{128 \lambda^2 \alpha  \zeta^4}{r^3}
			\right) \left(f^{\prime}\right)^2
		+	\left(
			-	\frac{512 \lambda^3 Q \alpha  \zeta^2}{r^4}
			-	\frac{256 \lambda^3 P \alpha  \zeta^3}{r^4}
			+	\frac{1024 \lambda^3 Q \alpha  \zeta^4}{r^4}
			+	\frac{768 \lambda^3 P \alpha  \zeta^5}{r^4}
			\right) \left(f^{\prime}\right)^3
		\right) \partial_r P
	\nonumber\\&
	+	\frac{
		256 \lambda^3 Q \alpha
		\zeta^3 \left(f^{\prime}\right)^3 \partial_r Q
		}{r^4}
	\Bigg) \partial_r\zeta
	\nonumber\\&
+	\left(
		\frac{
		64 \lambda^2 Q \alpha
		\zeta^2 \left(f^{\prime}\right)^2
		}{r^3}
	+	\left(
		-	\frac{512 \lambda^3 Q^2 \alpha  \zeta^2}{r^4}
		-	\frac{256 \lambda^3 P Q \alpha  \zeta^3}{r^4}
		\right) 
		\left(f^{\prime}\right)^3\right) \left(\partial_r\zeta\right)^2
	\nonumber\\&
+	\Bigg(
	-	Q
	-	P \zeta 
	+	\left(
			\frac{16 \lambda  Q^2}{r}
		+	\frac{40 \lambda  P Q \zeta }{r}
		-	\frac{8 \lambda  \zeta^2}{r^3}
		+	\frac{20 \lambda  P^2 \zeta^2}{r}
		+	\frac{4 \lambda  Q^2 \zeta^2}{r}
		+	\frac{16 \lambda  \zeta^4}{r^3}
		\right) f^{\prime}
	\nonumber\\&
	+	\Bigg(-\frac{64 \lambda^2 Q^3}{r^2}-\frac{256 \lambda^2
	P Q^2 \zeta }{r^2}+\frac{64 \lambda^2 Q \zeta^2}{r^4}
 	-\frac{256 \lambda^2 P^2 Q \zeta^2}{r^2}-\frac{32 \lambda^2 Q^3
	\zeta^2}{r^2}+\frac{64 \lambda^2 P \zeta^3}{r^4}
	-\frac{80 \lambda^2 P^3 \zeta^3}{r^2}
	\nonumber\\&
	-\frac{16 \lambda^2 P Q^2 \zeta^3}{r^2}
	-\frac{128 \lambda^2 Q \zeta^4}{r^4}
	-\frac{96 \lambda^2 P \zeta^5}{r^4}\Bigg) \left(f^{\prime}\right)^2
	\nonumber\\&
	+	\Bigg(\left(\frac{64 \lambda^2 Q^2 \zeta^2}{r^3}
	+\frac{192 \lambda^2 P Q \zeta^3}{r^3}
	+\frac{128 \lambda^2 P^2 \zeta^4}{r^3}\right) f^{\prime}
	\nonumber\\&
	+	\left(-\frac{512 \lambda^3 Q^3 \zeta^2}{r^4}
	-\frac{2048 \lambda^3 P Q^2 \zeta^3}{r^4}
	-\frac{2304 \lambda^3 P^2 Q \zeta^4}{r^4}
	-\frac{768 \lambda^3 P^3 \zeta^5}{r^4}\right)
	\left(f^{\prime}\right)^2\Bigg) f^{\prime\prime}
	\nonumber\\&
	+	\left(\left(\frac{192 \lambda^2 \zeta^3}{r^3}
	-\frac{128 \lambda^2 \zeta^5}{r^3}\right)
	\left(f^{\prime}\right)^2+\left(-\frac{1536 \lambda^3 Q \zeta^3}{r^4}
	-\frac{1280 \lambda^3 P \zeta^4}{r^4}
	+\frac{1024 \lambda^3 Q \zeta^5}{r^4}
	+\frac{768 \lambda^3 P \zeta^6}{r^4}\right)
	\left(f^{\prime}\right)^3\right) \partial_r P
	\nonumber\\&
	+	\left(\frac{64 \lambda^2 \zeta^2
	\left(f^{\prime}\right)^2}{r^3}+\left(
	-\frac{512 \lambda^3 Q \zeta^2}{r^4}
	-\frac{512 \lambda^3 P \zeta^3}{r^4}\right)
	\left(f^{\prime}\right)^3\right) \partial_r Q
	\nonumber\\&
	+	\Bigg(
		\frac{16 \lambda  \zeta  f^{\prime}}{r^2}
	+	\left(-\frac{256 \lambda^2 Q \zeta }{r^3}
	-\frac{192 \lambda^2 P \zeta^2}{r^3}
	+\frac{128 \lambda^2 Q \zeta^3}{r^3}\right) \left(f^{\prime}\right)^2
	\nonumber\\&
	+	\left(\frac{1024 \lambda^3 Q^2 \zeta }{r^4}
	+\frac{1536 \lambda^3 P Q \zeta^2}{r^4}
	+\frac{512 \lambda^3 P^2 \zeta^3}{r^4}
	-\frac{1024 \lambda^3 Q^2 \zeta^3}{r^4}
	-\frac{768 \lambda^3 P Q \zeta^4}{r^4}\right)
	\left(f^{\prime}\right)^3\Bigg) \partial_r\zeta
	\Bigg)
	\partial_r\alpha
	,\\
	&\mathcal{A}_{(\zeta)}
	=
	1
+	\left(-\frac{16 \lambda  Q}{r}-\frac{16 \lambda  P \zeta }{r}\right)
	f^{\prime}
+	\left(\frac{64 \lambda^2 Q^2}{r^2}+\frac{128 \lambda^2 P Q \zeta }{r^2}
	+\frac{64 \lambda^2 P^2 \zeta^2}{r^2}
	-\frac{32 \lambda^2 \zeta^4}{r^4}\right) \left(f^{\prime}\right)^2
	\nonumber\\&
+	\frac{256 \lambda^3 Q^2 \zeta^4 \left(f^{\prime}\right)^2
	f^{\prime\prime}}{r^4}
+	\frac{256 \lambda^3 \zeta^4 \left(f^{\prime}\right)^3 \partial_r Q}{r^4}
+	\left(\frac{128 \lambda^2 \zeta^4
	\left(f^{\prime}\right)^2}{r^3 \alpha }
	+\left(-\frac{1024 \lambda^3 Q \zeta^4}{r^4 \alpha }
	-\frac{768 \lambda^3 P \zeta^5}{r^4 \alpha }\right)
	\left(f^{\prime}\right)^3\right) \partial_r\alpha
	\nonumber\\&
+	\left(\frac{128 \lambda^2 \zeta^3 \left(f^{\prime}\right)^2}{r^3}
	+\left(-\frac{1024 \lambda^3 Q \zeta^3}{r^4}
	-\frac{768 \lambda^3 P \zeta^4}{r^4}\right)
	\left(f^{\prime}\right)^3\right) \partial_r\zeta
	,\\
	&\mathcal{F}_{(\zeta)}
	=	 
-	\frac{1}{4} r P^2 \alpha 
-	\frac{1}{4} r Q^2 \alpha 
-	\frac{r P Q \alpha }{2 \zeta }
-	\frac{\alpha  \zeta^2}{2 r}
	\nonumber\\&
+	\left(6 \lambda  P^2 Q \alpha +2 \lambda  Q^3 \alpha
	+\frac{4 \lambda  P Q^2 \alpha }{\zeta }
	+2 \lambda  P^3 \alpha  \zeta 
	+2 \lambda  P Q^2 \alpha  \zeta 
	-\frac{4 \lambda  Q \alpha  \zeta^2}{r^2}
	-\frac{4 \lambda  P \alpha  \zeta^3}{r^2}\right) f^{\prime}
	\nonumber\\&
+	\left(\frac{64 \lambda^2 Q^2 \alpha  \zeta^2}{r^3}
	+\frac{128 \lambda^2 P Q \alpha  \zeta^3}{r^3}
	+\frac{80 \lambda^2 P^2 \alpha  \zeta^4}{r^3}
	-\frac{16 \lambda^2 Q^2 \alpha  \zeta^4}{r^3}
	\right) \left(f^{\prime}\right)^2
	\nonumber\\&
+	\left(-\frac{4 \lambda  P Q \alpha  \zeta }{r}
	-\frac{4 \lambda  P^2 \alpha  \zeta^2}{r}
	+\left(\frac{32 \lambda^2 P Q^2 \alpha  \zeta }{r^2}
	+\frac{64 \lambda^2 P^2 Q \alpha  \zeta^2}{r^2}
	+\frac{32 \lambda^2 P^3 \alpha  \zeta^3}{r^2}
	\right) f^{\prime}\right) f^{\prime\prime}
	\nonumber\\&
+	\left(-\frac{4 \lambda  \alpha  \zeta  f^{\prime}}{r}
	+\left(\frac{32 \lambda^2 Q \alpha  \zeta }{r^2}
	+\frac{32 \lambda^2 P \alpha  \zeta^2}{r^2}\right)
	\left(f^{\prime}\right)^2\right) \partial_r P
	\nonumber\\&
+	\left(-\frac{4 \lambda  \alpha  \zeta^2 f^{\prime}}{r}
	+\left(\frac{32 \lambda^2 Q \alpha  \zeta^2}{r^2}
	+\frac{32 \lambda^2 P \alpha  \zeta^3}{r^2}\right)
	\left(f^{\prime}\right)^2\right) \partial_r Q
	\nonumber\\&
+	\left(\frac{256 \lambda^3 P \zeta^5}{r^4 \alpha }
	+\frac{256 \lambda^3 Q \zeta^6}{r^4 \alpha }\right)
	\left(f^{\prime}\right)^3 \left(\partial_r\alpha\right)^2
	\nonumber\\&
+	\Bigg(
	-	\alpha  \zeta 
	+	\left(\frac{12 \lambda  Q \alpha  \zeta }{r}
	+\frac{12 \lambda  P \alpha  \zeta^2}{r}\right) f^{\prime}
	\nonumber\\&
	+	\left(-\frac{32 \lambda^2 Q^2 \alpha  \zeta }{r^2}
	-\frac{96 \lambda^2 P Q \alpha  \zeta^2}{r^2}
	-\frac{48 \lambda^2 P^2 \alpha  \zeta^3}{r^2}
	-\frac{16 \lambda^2 Q^2 \alpha  \zeta^3}{r^2}
	+\frac{32 \lambda^2 \alpha  \zeta^5}{r^4}\right) 
	\left(f^{\prime}\right)^2
	\nonumber\\&
	+	\left(-\frac{256 \lambda^3 P Q \alpha  \zeta^4}{r^4}
	-\frac{256 \lambda^3 Q^2 \alpha  \zeta^5}{r^4}\right)
	\left(f^{\prime}\right)^2 f^{\prime\prime}
	-\frac{256 \lambda^3 \alpha  \zeta^4 
	\left(f^{\prime}\right)^3 \partial_r P}{r^4}
	-\frac{256 \lambda^3 \alpha  \zeta^5
	\left(f^{\prime}\right)^3 \partial_r Q}{r^4}\Bigg) \partial_r\zeta
	\nonumber\\&
+	\left(-\frac{128 \lambda^2 \alpha  \zeta^4
	\left(f^{\prime}\right)^2}{r^3}
	+\left(\frac{768 \lambda^3 Q \alpha  \zeta^4}{r^4}
	+\frac{768 \lambda^3 P \alpha  \zeta^5}{r^4}\right)
	\left(f^{\prime}\right)^3\right) \left(\partial_r\zeta\right)^2
	\nonumber\\&
+	\Bigg(
	\left(-\frac{4 \lambda  Q \zeta^2}{r}
	-\frac{4 \lambda  P \zeta^3}{r}\right) f^{\prime}
	+\left(\frac{32 \lambda^2 Q^2 \zeta^2}{r^2}
	+\frac{32 \lambda^2 P Q \zeta^3}{r^2}-\frac{32 \lambda^2 \zeta^4}{r^4}
	+\frac{16 \lambda^2 P^2 \zeta^4}{r^2}
	-\frac{16 \lambda^2 Q^2 \zeta^4}{r^2}\right) \left(f^{\prime}\right)^2
	\nonumber\\&
	+\left(\frac{256 \lambda^3 Q^2 \zeta^4}{r^4}
	+\frac{256 \lambda^3 P Q \zeta^5}{r^4}\right)
	\left(f^{\prime}\right)^2 f^{\prime\prime}
	+\frac{256 \lambda^3 \zeta^5
	\left(f^{\prime}\right)^3 \partial_r P}{r^4}
	+\frac{256 \lambda^3 \zeta^4
	\left(f^{\prime}\right)^3 \partial_r Q}{r^4}
	\nonumber\\&
	+\left(\left(\frac{64 \lambda^2 \zeta^3}{r^3}
	-\frac{128 \lambda^2 \zeta^5}{r^3}\right)
	\left(f^{\prime}\right)^2
	+\left(-\frac{512 \lambda^3 Q \zeta^3}{r^4}
	-\frac{256 \lambda^3 P \zeta^4}{r^4}
	+\frac{1024 \lambda^3 Q \zeta^5}{r^4}
	+\frac{768 \lambda^3 P \zeta^6}{r^4}\right)
	\left(f^{\prime}\right)^3\right) \partial_r\zeta\Bigg)
	\partial_r\alpha 
	.
\end{align}
\end{subequations}
%%%%%%%%%%%%%%%%%%%%%%%%%%%%%%%%%%%%%%%%%%%%%%%%%%%%%%%%%%%%%%%%%%%%%%%%%%%%%%
\bibliography{globalbib}

\end{document}